\documentclass[journal,onecolumn,11pt]{IEEEtran}
\usepackage{cite}
\usepackage{amsbsy} 
\usepackage{amsmath}
\usepackage{booktabs}
\usepackage{mathtools}
\usepackage{amsfonts} 
\usepackage{amssymb}  
\usepackage{amsthm}   
\usepackage[ruled]{algorithm2e}
\usepackage{algorithmic}
\usepackage{amsxtra} 
\usepackage{bm}
\usepackage{multirow}
\usepackage{nomencl}
\usepackage{color,colortbl}
\usepackage{dsfont}
\usepackage{enumerate}
\usepackage{enumitem}
\usepackage{fancybox}
\usepackage{fancyhdr}
\usepackage{float}
\usepackage{graphics}
\usepackage{graphicx} 
\usepackage{latexsym}
\usepackage{longtable}
\usepackage{subfig}
\usepackage{url}
\def\BState{\State\hskip-\ALG@thistlm}
\usepackage[utf8]{inputenc}
\usepackage[english]{babel}

\usepackage[ruled]{algorithm2e}
\usepackage{algorithmic}
\usepackage{amsfonts}
\usepackage{amssymb}
\usepackage{float}
\usepackage{multicol}
\usepackage{multirow}

\DeclareMathOperator{\Tr}{Tr}

\usepackage{tikz}

\theoremstyle{remark}     
\newtheorem*{remark*}{Remark}   

\definecolor{Dblue}{rgb}{0,0,1}
\definecolor{Dbrown}{rgb}{0.59,0.4,0}
\definecolor{Dred}{rgb}{0.64,0,0}
\definecolor{Dgreen}{rgb}{0,0.4,0}

\newcommand*{\qeda}{\hfill\ensuremath{\blacksquare}}

\begin{document}
	
\title{Using growth transform dynamical systems for spatio-temporal data sonification}
\author{Oindrila~Chatterjee,~\IEEEmembership{Student~Member,~IEEE}, and~Shantanu~Chakrabartty,~\IEEEmembership{Senior~Member,~IEEE}\\	
	{Washington University in St. Louis, Missouri 63130, USA. }\\
	\thanks{All correspondences regarding this manuscript should be addressed to shantanu@wustl.edu.}
	\thanks{This work was supported in part by a research grant from the National Science Foundation (ECCS:1935073).}}
\maketitle

\section{Abstract}
\label{cpt:sonification}
Sonification, or encoding information in meaningful audio signatures, has several advantages in augmenting or replacing traditional visualization methods for human-in-the-loop decision-making. Standard sonification methods reported in the literature involve either (i) using only a subset of the variables, or (ii) first solving a learning task on the data and then mapping the output to an audio waveform, which is utilized by the end-user to make a decision. This paper presents a novel framework for sonifying high-dimensional data using a complex growth transform dynamical system model where both the learning (or, more generally, optimization) and the sonification processes are integrated together. Our algorithm takes as input the data and optimization parameters underlying the learning or prediction task and combines it with the psychoacoustic parameters defined by the user. As a result, the proposed framework outputs binaural audio signatures that not only encode some statistical properties of the high-dimensional data but also reveal the underlying complexity of the optimization/learning process.  
Along with extensive experiments using synthetic datasets, we demonstrate the framework on sonifying Electro-encephalogram (EEG) data with the potential for detecting epileptic seizures in pediatric patients. 
\section{Introduction}
With the multitude of high-dimensional data available today, the search for better techniques for perceptualizing data before applying it to a learning/predictive task has emerged as an important research area. Though visualization remains the modality of choice for most applications, sonification (representation of data using human-recognizable audio signatures \cite{hermann2011sonification}) is gaining momentum as a complementary or alternative modality.
\par Some of the advantages of data sonification, that make it an ideal candidate for augmenting or replacing visualization for analyzing the dataset's properties, are as follows \cite{hermann2011sonification,kramer2010sonification,dubus2013systematic}:
\begin{enumerate}
	\item Temporal resolution for auditory perception is better than that for visual perception. This makes it suitable for perceptualizing time-varying data with complex patterns that visual displays might otherwise miss.
	\item Audio is orientation-agnostic and has a wider spatial range since the user need not be oriented towards a particular direction. In contrast, for visualization, objects need to be within the field of vision of the user.
	\item Humans typically respond faster to auditory feedback when compared to visual feedback, making sonification attractive for human-in-the-loop control applications.
	\item In scenarios where the visual scene is crowded due to multiple displays or when attention is lacking, auditory signals can be used as a means of drawing the user's attention to a particular segment of the visual field.
	\item Auditory perception provides a natural alternative to shrinking display sizes, especially for monitoring and alerting applications.  
	\item Auditory perception and auditory memory remain resilient to many neurodegenerative diseases, as a result sonification is an attractive choice for rehabilitation technologies.
\end{enumerate} 

\par
\begin{figure}[!h]
	\begin{center}
		\includegraphics[scale=0.7]{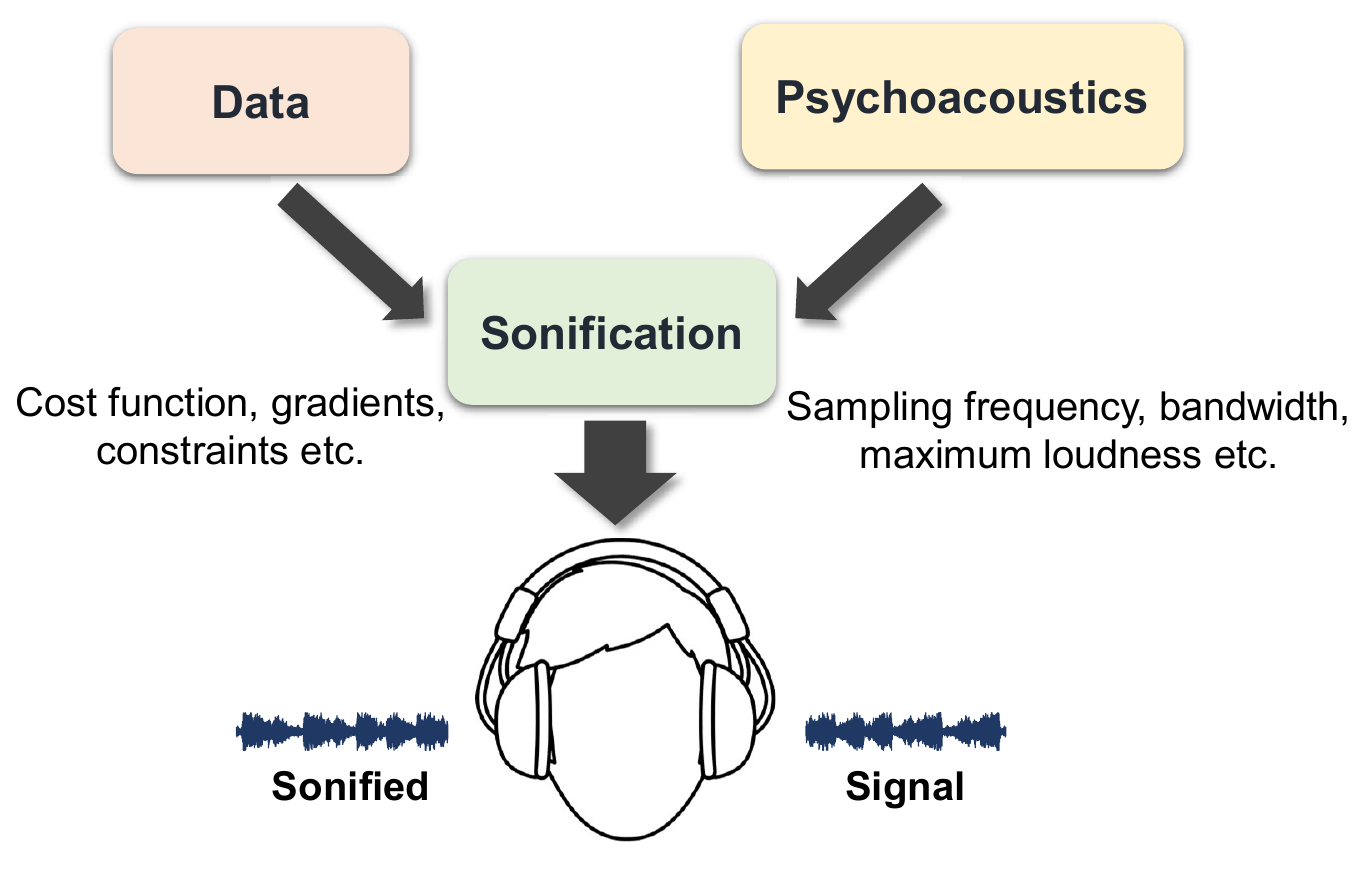}
		\caption[Proposed approach for data sonification]{\textbf{ Proposed approach for data sonification.}
			The proposed sonification module takes as input (i) a generic optimization/learning problem defined by a cost function, gradients, and constraints on one hand, and (ii) psychoacoustic parameters like sampling frequency  etc. on the other. It then outputs a single or dual-channel audio waveform that encodes the underlying optimization problem.}
		\label{fig1}
	\end{center}
\end{figure}

 The most common sonification scheme used in literature involves mapping only a subset of the most important variables in the data directly within the audible range \cite{kather2017polyphonic,loui2014rapidly}. A second strategy involves using a machine learning algorithm to perform a classification task and then sonifying the output of the classifier \cite{zhou2018epileptic}. While the first approach does not utilize the entire information available at hand, the second approach lacks a human-in-the-loop component that might be crucial for tasks involving monitoring and feedback. In this paper, we propose a novel sonification technique that incorporates both the learning task at hand, as well as psychoacoustic parameters defined by the end-user in the form of constraints placed on the task. The algorithm finally produces an audio signature that can be utilized by the end-user for identifying both spatial and temporal patterns in the dataset. The scope of the paper is illustrated in Fig~\ref{fig1}. At the core of the approach is a variant of the complex growth transform dynamical system, proposed in our previous work \cite{chatterjee2018decentralized, chatterjee2020resonant} based on Baum-Eagon growth transforms \cite{baum1968growth}. Effectively, the high-dimensional space encompassed by the learning task is projected into a network of interacting limit cycle oscillators that form a set of complex basis functions encoding the low-dimensional space of the audio signal. The main model along with its properties is presented in Section \ref{sec_complexGT}, while Section \ref{sec_H_effect} shows the effects of the learning task on the sonified signal. Section \ref{sec_sonification_methods} presents the effects of the psychoacoustic parameters and sonification strategies that can be adopted for the sonification module. Section \ref{sec_syn_exp} presents the application of the sonification framework on synthetic datasets to highlight the role of different parameters and sonification strategies in the process. Section \ref{sec_real_exp}, finally, shows how our model can be applied to the CHB-MIT scalp EEG dataset \cite{shoeb2010application,goldberger2000physiobank} for the real-time detection of epileptic seizures.
\par In summary, the key contributions of the paper can be highlighted as follows:
\begin{enumerate}
	\item The proposed method inherently maps high-dimensional data into a lower-dimensional single-channel or dual-channel audio signal. This is particularly suitable when dealing with low throughput systems having limited capacity/ bandwidth.
	\item The method utilizes the whole spectrum of information available for a task and enables the end-user to arrive at a decision instead of relying on a machine learning model. 
	\item The algorithm automatically combines the learning and sonification stages into a single module. This is unlike the sonification-based decision-making algorithms existing in literature, that typically involve learning the decision parameters and then mapping the same to different sound parameters in subsequent stages.
	\item The method is versatile and can be tailored to accommodate various learning models, problem dimensionality, and dataset sizes. Additionally, the complex growth transform dynamical system provides a wide range of tunable parameters which can be customized for different applications.
\end{enumerate}

\section{Background and Related Work}
\label{sec_background}

\subsubsection{Visualization of High-Dimensional Data}
\label{subsec_visualizationlit}

\par Elementary visualization schemes use different visual attributes like color, shape, size, spatial location, etc., for representing information. However, almost all the standard visualization techniques typically use these attributes after mapping the data to a lower-dimensional (usually two or three-dimensional) space. Some of the most commonly used mapping techniques include linear methods like PCA, LDA and their variants \cite{martinez2001pca} that only preserve the global structure of the data. In contrast, nonlinear methods like Locally Linear Embedding, \cite{roweis2000nonlinear}, Isomap \cite{balasubramanian2002isomap}, Laplacian EigenMap \cite{belkin2003laplacian}, tSNE \cite{maaten2008visualizing}, PHATE \cite{moon2019visualizing} etc, have been designed to preserve both global and local information. 
\par However, these techniques are more suitable for time-invariant or static data, and analysis of time-varying, high-dimensional data using these methods would involve frame-wise visualization in a lower-dimensional space. Several visualization methods based on variants of the techniques discussed earlier have been proposed, e.g., m-tSNE \cite{nguyen2017m}. However, they lack interpretability and suffer from a similar limitation of being restricted to only three dimensions for representing the data. Moreover, though other visual attributes like color, size, shape, etc., can be used in conjunction with the three spatial dimensions, these may not be sufficient for complex datasets with very high dimensionality. This can thus potentially lead to information overload of the visual system, thereby pushing the users to the limits of comprehension. Additionally, complex, time-varying patterns in the data are sometimes difficult to capture using only the visual faculties. Hence, employing sonification might help users interpret the data more effectively or draw their attention to the most important aspects of the data that can then be monitored/analyzed by visual inspection \cite{hermann2011sonification}.

\subsubsection{Data Sonification and its Applications}
\label{subsec_sonificationlit}

\par Sonification, similar to its visual counterpart, provides several attributes that can be used for representing information, like pitch, loudness, timbre, rhythm, duration, harmonic content, etc. Broadly, two different sonification methods exist in the literature depending on how the data is mapped into an audio signal: (i) parameter mapping sonification and (ii) model-based sonification \cite{hermann2011sonification}. Of these, the parameter mapping method, where variables can be mapped into different attributes of the sound signal to create a sonified signature, is the most commonly used. Parameter mapping based sonification has been used in a wide range of applications, ranging from sonifying astronomical data like gravitational waves \cite{st2018sonification} and photons emitted by the Higgs bosons \cite{supper2014sublime}, to synthesizing novel protein structures \cite{dunn1999life,yu2020sonification} and detecting anomalies in medical data (e.g., CT scans of Alzheimer's patients \cite{gionfrida2017novel}, EEG \cite{valjamae2013review} and ECG \cite{kather2017polyphonic} signals, skin cancer detection \cite{dascalu2019skin}, epileptic seizure detection \cite{loui2014rapidly}), detection of anomalous events \cite{ballora2012use} etc. Sonification based techniques have also been used for intrusion detection in networks \cite{axon2017formalised} and network traffic flow \cite{debashi2018sonification}, analysis of stock market data \cite{janata2004marketbuzz}, and in therapeutic treatment of freezing of gait in Parkinson's patients \cite{mezzarobba2018action}. More recently, sonification has been applied to the analysis of RNA sequences in different strains of the Covid19 virus \cite{buehler2020nanomechanical}. Unlike the visualization schemes, sonification offers a much wider parameter space for the variables to be mapped into. For example, humans can perceptualize sound signals anywhere from 20 to 20kHz, and frequency differences as low as 3 Hz are easily discernible by the human auditory system \cite{hermann2011sonification}. In contrast, the model-based sonification technique involves using a virtual model whose sonic responses are altered according to the data provided. This might involve first passing the data through a machine learning or optimization module which learns the manifold the data resides in, and then mapping the output to different properties of the sound signal. While the parameter mapping based method is restricted in terms of the dimensionality of the data it can handle, the model-based approach uses the entire data available, but does not rely on the end user for the decision-making \cite{hermann2011sonification}. 

\section{Sonification Framework Using Complex Growth Transforms}
\label{sec_complexGT}

In this section, we introduce a novel framework for sonifying high-dimensional data using a variant of the complex domain dynamical system model proposed in our previous work\cite{chatterjee2020resonant}. Our method utilizes the entire information available at hand to perform a combination of feature extraction and dimensionality reduction. Finally, it allows the human user to arrive at a decision based on the output sound signature. The sonification algorithm has been summarized in Table \ref{son_algo1}, while the proof has been outlined in \ref{S1_appendix}. The inputs to the sonification process are: (i) the learning parameters and data under consideration; and (ii) the psychoacoustic parameters defined by the user. We then develop a time-evolution operator $\mathcal{U}$ that depends on both the learning task and the psychoacoustics by incorporating the latter as additional constraints on the learning/optimization process. $\mathcal{U}$ is chosen to be a nonlinear unitary transformation that ensures that the total energy of the sonified signal remains conserved over time. This effect mimics an automatic gain control mechanism observed in bioacoustics\cite{lyon1990automatic}. 

\par
\begin{figure}[!h]
	\begin{center}
		\includegraphics[scale=0.6, trim=2 2 2 2,clip]{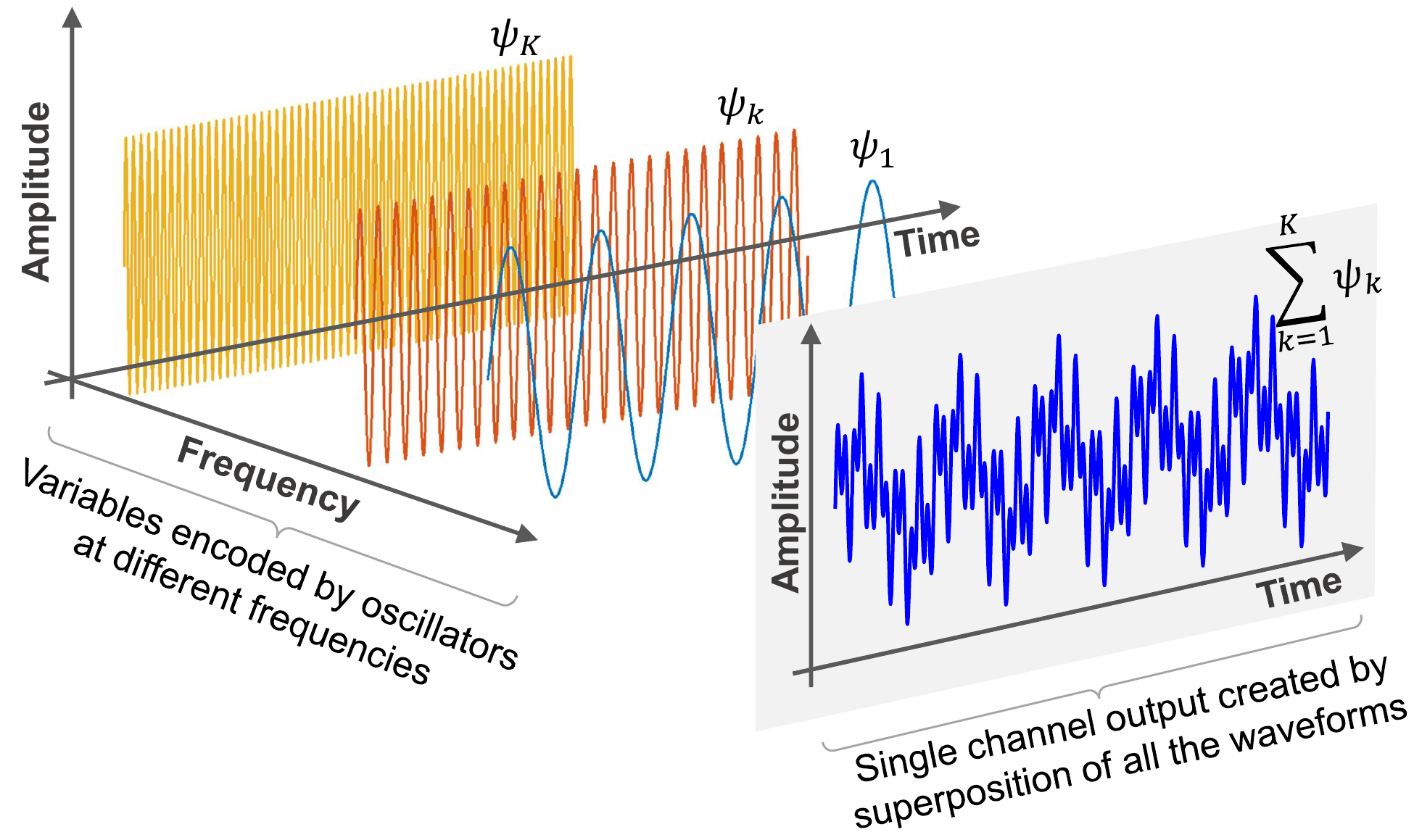}
		\caption[Illustration of the sonification process]{\textbf{Illustration of the sonification process.}
			$K$ oscillator waveforms $\psi_1,\ldots,\psi_K$, representing the set of complex basis functions defined by the user, based on psychoacoustics. Both amplitude and frequency of the basis set oscillators get modulated over time based on the learning task. The sonified output is created by the superimposition of all the oscillator signals, i.e., $\psi_{\text{sum}(t) }=\sum\limits_{k=1}^K \psi_k(t)$.}
		\label{fig2}
	\end{center}
\end{figure}

\begin{figure}[!h]
	\begin{center}
		\includegraphics[scale=0.7, trim=2 2 2 2,clip]{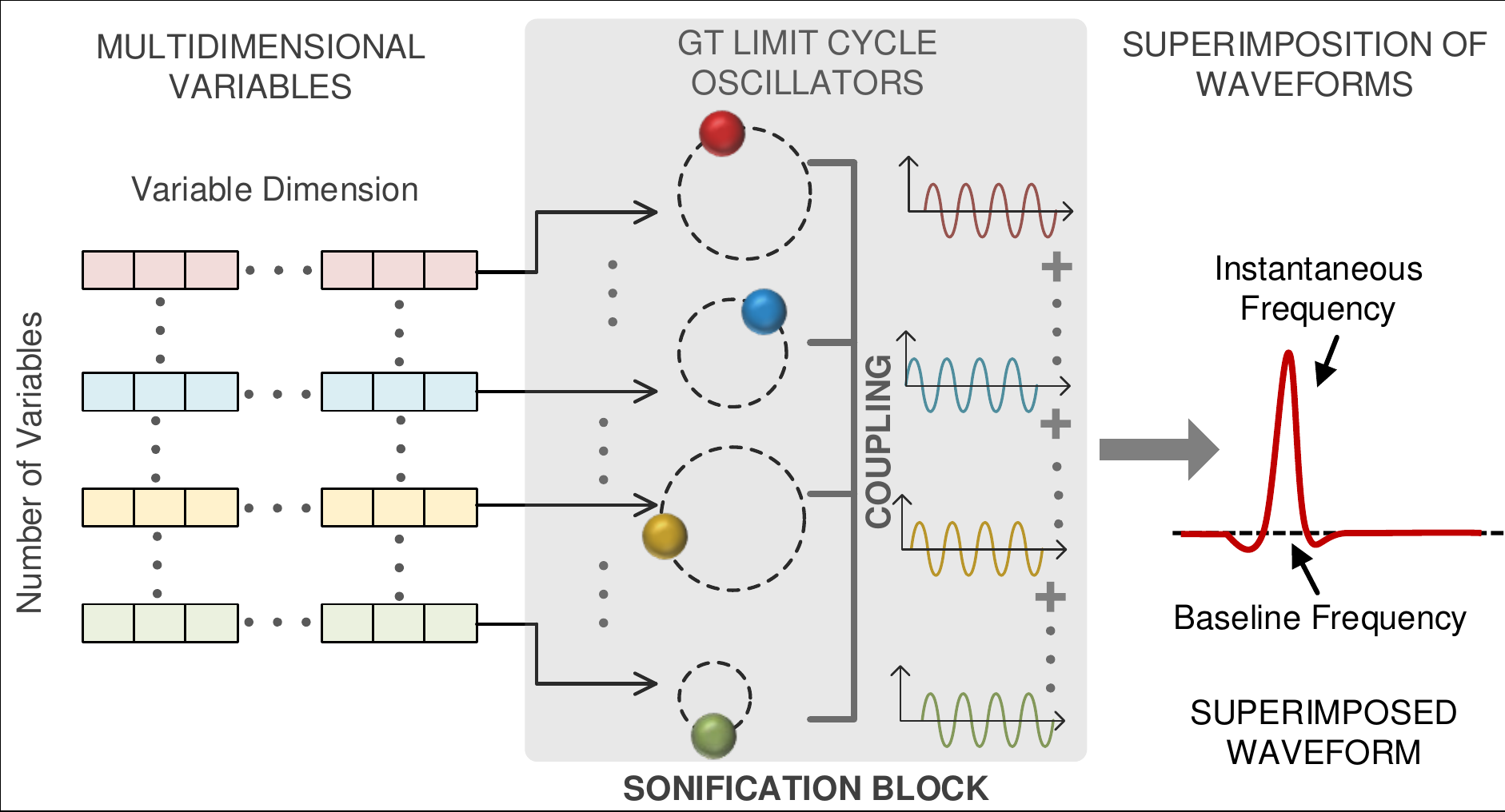}
		\caption[Superimposition of oscillators]{\textbf{ Illustration of the superimposition of oscillators.}
			Each multidimensional variable/data vector is mapped into an individual growth transform limit cycle oscillator. Local and global couplings lead to a constant phase difference between each pair of oscillators in a steady-state. This results in a sudden shift of the frequency of the superimposed waveform $\psi_{\text{sum}}(t)$ from the baseline frequency during the transient state, following which the frequency is restored to its baseline value in the steady-state.}
		\label{fig3}
	\end{center}
\end{figure}

 Our proposed technique is thus a blend of both the parameter mapping-based and model-based sonification strategies. This is because the entire range of the data is mapped to different parameters of the output sound signal based on a dynamical system model. However, the end-user takes the decision, as illustrated in Fig~\ref{fig1}. Within the framework, each variable or data point is mapped to a set of complex growth transform limit cycle oscillators globally coupled together by the conservation constraint on the signal energy. The oscillators can be thought of as complex basis functions defining the lower-dimensional space of the audio output to which we want to project our high-dimensional data, and are dictated by the psychoacoustic parameters. This is illustrated in Fig~\ref{fig2} for a group of $K$ oscillators represented by complex oscillator variables or waveforms $\psi_1(t),\ldots, \psi_K(t) $ having different frequencies and amplitudes. The variables involved are assigned both baseline and relative frequencies as outlined in Appendix \ref{S1_appendix}. As a result of the time evolution of the complex dynamical system, both the amplitudes and frequencies of the oscillators get modulated during the optimization process. For example, if all the variables get mapped to the same constant value of baseline and relative frequencies, each oscillator frequency drifts from this constant value during the transient phase, following which it again returns to the original trajectory. The frequency deviation during the transient phase should ideally encode the complexity of the optimization problem. The output sonified signal is thus a complex single or dual-channel waveform obtained by a superposition of all the oscillator waveforms, ie., $\psi_{\text{sum} }(t)=\sum\limits_{k=1}^K \psi_k(t)$, as. illustrated in Fig~\ref{fig3}.

\subsection{Effect of the learning problem on sonified output} \label{sec_H_effect}

\par In this section, we will show how the learning problem affects the sonified signal. Any statistical learning task can be expressed in terms of an objective function $\mathcal{H}$ that we are trying to minimize over a constraint space. We will therefore investigate the role of the objective function on the sound signature. Considering default temporal constraints on the total signal energy, the complex growth transform-based sonification technique (please see Table \ref{son_algo1}) provides a natural framework for optimizing the cost function $\mathcal{H}$ subject to constraints defined by the psychoacoustics. Without loss of generality, in this section, we will focus only on the effect of $\mathcal{H}$ and keep the psychoacoustic parameters constant for all the experiments. In particular, we will consider here that (a) the total energy of the sonified signal remains conserved over time, and  (b) all the oscillators or basis waveforms in the network have identical baseline frequencies for better visualization of the network dynamics. The sonified output signal for all cases, as stated earlier, is obtained by the superposition of the wave functions generated by all the oscillators, i.e., $\Psi_{\text{sum},n}=\sum \limits_{k=1}^K \Psi_{k,n}$.
\setcounter{algocf}{1}
\begin{algorithm*}[!ht]
	\caption{Sonification using complex growth transform dynamical system (Proof in Appendix \ref{S1_appendix})} 
	\label{son_algo1}
	\begin{algorithmic}
		\vspace{0.2cm} 
		\STATE $\bullet$ \textbf{Learning task:} Consider a statistical learning task defined by a cost function and gradients. 
		\vspace{0.1cm}
		\STATE $\bullet$ \textbf{Psychoacoustics:} Consider  a set of complex basis functions $\boldsymbol{\mathbf{\Psi}}=[\psi_1,\ldots,\psi_K] \in \mathbb{C}^K$ defined by the user using psychoacoustic parameters like sampling frequency $Fs$, bandwidth ($f_1$, $f_2$), frequency mapping strategy etc. 
		\vspace{0.1cm}
		\STATE $\bullet$ \textbf{Sonification module:} The growth transform-based sonification method produces an output signal of the following form in steady-state:
		\begin{equation}
			\psi_{\text{sum},n} = \sum \limits_{k=1}^K \psi_{k,n} = \sum \limits_{k=1}^K \psi_{k,n-1} \exp \Bigg(j(\omega_{n}+ \xi_{k,n})\dfrac{1}{Fs}\Bigg),	
		\end{equation}
		where $\omega_{n}$ and $\xi_{k,n}$ are the instantaneous baseline and relative frequencies of the $k^\text{th}$ oscillator variable $\psi_{k,n} \in \mathbb{C}$ at the $n^\text{th}$ time step respectively.
		\\
		\vspace{0.1cm} 
		\STATE $\bullet$ \textbf{Update rule:} The corresponding instantaneous time evolution equation for the oscillators is given by the following equation:
		\begin{equation}
			\boldsymbol{\Psi}_n \leftarrow \mathcal{U}_{n}(\boldsymbol{\Psi}_{n-1}) \odot \boldsymbol{\Psi}_{n-1},
		\end{equation}
		where the mathematical form of $\mathcal{U}_{n}$ is given in Appendix \ref{S1_appendix}.
		\\
		\vspace{0.1cm}
		\STATE $\bullet$ \textbf{Conservation of signal power:} The operator $\mathcal{U}_{n}: \mathbb{C}^K \rightarrow \mathbb{C}^K$ represents an instantaneous nonlinear unitary transformation that ensures that the total signal energy remains conserved over time and thus imposes a temporal constraint brought about by the psychoacoustics as well, i.e.,
		\begin{equation}
			\sum \limits_{k=1}^K \lvert\psi_{k,n}\rvert^2 = \gamma \quad \forall n, \quad \gamma > 0.	
		\end{equation}
		\vspace{0.2cm}
	\end{algorithmic}
\end{algorithm*}
\newline \textbf{Example 1:} Consider the following one-dimensional quadratic optimization problem:
\begin{gather}
	\underset{p \in \mathbb{R} } {\text{minimize}} \quad  \mathcal{H}_1^{\prime}(p)=8p^2-2p \nonumber \label{eq_ex1}\\
	\textit{s.t.}\quad \lvert p \rvert \le \gamma 
\end{gather}

\begin{figure}[!h]
	\begin{center}
		\includegraphics[scale=0.6]{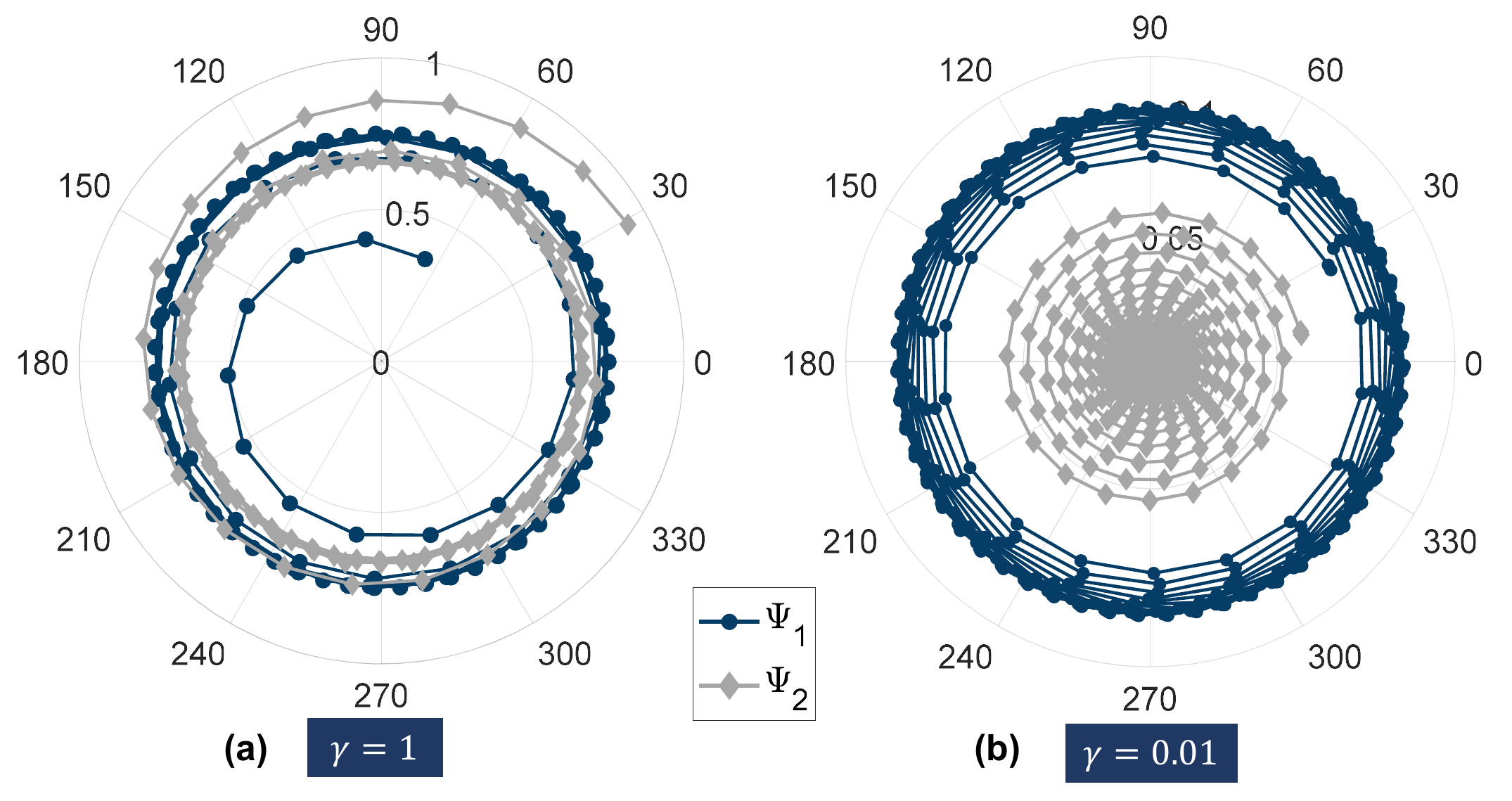}
		\caption[Effect of $\gamma$]{\textbf{ Effect of $\gamma$.}
			Dynamics of growth transform-based complex dynamical system consisting of two oscillators for two different cases corresponding to problem $\mathcal{H}_1$: (a) both oscillators asymptotically reach their respective stable limit cycles in steady-state ($\gamma=1$) and (b) one reaches a stable limit cycle, while the other goes to a stationary point ($\gamma=0.01$).}
		\label{fig4}
	\end{center}
\end{figure}

We can map the above to the sonification framework in Table \ref{son_algo1} by taking $p=\lvert \psi_1 \rvert ^2 -\lvert \psi_2 \rvert ^2$, where $\psi_1, \psi_2 \in \mathbb{C}$ are complex waveforms. (Please see Appendix \ref{S1_appendix} for a detailed outline of the mapping procedure.) The equivalent optimization problem in the complex domain thus becomes:
\begin{gather}
	\underset{\boldsymbol{\Psi} \in \mathbb{C}^2 } {\text{minimize}} \quad  \mathcal{H}_1(\boldsymbol{\Psi})=8(\lvert \psi_1 \rvert ^2 -\lvert \psi_2 \rvert ^2)^2-2(\lvert \psi_1 \rvert ^2 -\lvert \psi_2 \rvert ^2) \nonumber \label{eq_ex1_map}\\
	\textit{s.t.}\quad \lvert \psi_1 \rvert ^2 +\lvert \psi_2 \rvert ^2 = \gamma, 
\end{gather}
where $\boldsymbol{\Psi}=[\psi_1,\psi_2]$. For this simple problem, the expression for the unitary operator $\mathcal{U}$ using the procedure outlined in Table \ref{son_algo1} (please see Appendix \ref{S1_appendix} for a detailed derivation) is given by:
\begin{gather}
	\mathcal{U}_{n} = \begin{pmatrix}
		d_{1,n}   \\
		d_{2,n} 
	\end{pmatrix},\nonumber \\
	d_{k,n}=\Big[\cos\Big(\dfrac{\omega_{n}}{Fs}\Big) 
	+j\sin \Big(\dfrac{\omega_{n}}{Fs} \Big)\sigma_{k,n-1} \Big]\exp \Big(j\dfrac{\xi_{k,n}}{Fs}\Big ), 	\nonumber \\
	\sigma_{k,n-1} =\sqrt{\dfrac{\Big(-\dfrac{\partial H}{\partial  \psi_{k,n-1}}+\lambda \psi_{k,n-1}^*\Big)}{\psi_{k,n-1}^*\sum\limits_{l=1}^{K=2} \psi_{l,n-1}\Big(-\dfrac{\partial \mathcal{H}}{\partial \psi_{l,n-1}}+\lambda \psi_{l,n-1}^*\Big)}}, \quad k=1,2.
\end{gather}
Fig \ref{fig4} shows the polar coordinate evolution plots of the waveforms $\psi_1$ and $\psi_2$ respectively, for two different choices of the parameter $\gamma$: (a) $\gamma=1$ and (b) $\gamma=0.01$. It can be seen that the oscillators corresponding to $\psi_1$ and $\psi_2$ converge to steady limit cycle oscillations for larger values of $\gamma$. However, for very small $\gamma$ values, only the oscillator with the maximum amplitude shows sustained oscillations, while the other converges to the fixed point of zero and stops oscillating. The coupled oscillator network can thus be thought of as some form of a frequency tuner for sufficiently small $\gamma$.

\textbf{Example 2:} Consider a second quadratic optimization problem:
\begin{gather}
	\underset{\mathbf{p} \in \mathbb{R}^M} {\text{minimize}} \quad  \mathcal{H}_2^{\prime}(\mathbf{p})=\dfrac{1}{2}a\mathbf{p^T Q p}-\mathbf{c^Tp} \nonumber \label{eq_ex2}\\
	\textit{s.t.}\quad \lvert p_i \rvert \le \gamma \enskip \forall i=1,\ldots,M, \quad  \gamma \in \mathbb{R}_+,
\end{gather}
and $\mathbf{c}\in \mathbb{R}^M, \mathbf{Q}\in \mathbb{R}^{M \times M}$. $a \in \mathbb{R}_+$ is a parameter that controls the curvature of the optimization surface, and hence controls the rate of convergence to the optimal solution. Taking $p_i=\lvert \psi_{i1} \rvert ^2 -\lvert \psi_{i2} \rvert ^2$, where $\psi_{i1}, \psi_{i2} \in \mathbb{C}$, we can again follow the mapping procedure in Appendix \ref{S2_appendix}. The equivalent optimization problem in the complex domain is given by:
\begin{gather}
	\underset{\boldsymbol{\Psi}_{1},\boldsymbol{\Psi}_{2} \in \mathbb{C}^M} {\text{minimize}} \quad  \mathcal{H}_2(\boldsymbol{\Psi}_{1},\boldsymbol{\Psi}_{2})=\dfrac{1}{2}a(\lvert \boldsymbol{\Psi}_{1} \rvert ^2 -\lvert \boldsymbol{\Psi}_{2} \rvert ^2)^T \mathbf{Q} (\lvert \boldsymbol{\Psi}_{1} \rvert ^2 -\lvert \boldsymbol{\Psi}_{2} \rvert ^2) \\
	-\mathbf{c^T}(\lvert \boldsymbol{\Psi}_{1} \rvert ^2 -\lvert \boldsymbol{\Psi}_{2} \rvert ^2) \nonumber \label{eq_ex2_map}\\
	\textit{s.t.}\quad \lvert \psi_{i1} \rvert ^2 +\lvert \psi_{i2} \rvert ^2 =  \gamma \enskip \forall i=1,\ldots,M, \quad  \gamma \in \mathbb{R}_+,
\end{gather}
where $\boldsymbol{\Psi}_1=\{\psi_{11},\ldots,\psi_{i1},\ldots, \psi_{M1}\}$ and $\boldsymbol{\Psi}_2=\{\psi_{12},\ldots,\psi_{i2},\ldots,\psi_{M2}\}$. Again, defining the unitary operator $\mathcal{U}$ according to Appendix \ref{S1_appendix}, we have:
\begin{gather}
	\mathcal{U}_{n} = \begin{pmatrix}
		d_{11,n} & d_{12,n}   \\
		\vdots & \vdots \\
		d_{i1,n} & d_{i2,n} \\
		\vdots & \vdots \\
		d_{M1,n} & d_{M2,n}
	\end{pmatrix}, \quad \mathcal{U}_n : \mathbb{C}^{M\times 2} \mapsto \mathbb{C}^{M\times 2}\nonumber \\
	d_{ik,n}=\Big[\cos\Big(\dfrac{\omega_{i,n}}{Fs}\Big) 
	+j\sin \Big(\dfrac{\omega_{i,n}}{Fs} \Big)\sigma_{ik,n-1} \Big]\exp \Big(j\dfrac{\xi_{ik,n}}{Fs}\Big ), 	\nonumber \\
	\sigma_{ik,n-1} =\sqrt{\dfrac{\Big(-\dfrac{\partial H}{\partial  \psi_{ik,n-1}}+\lambda \psi_{ik,n-1}^*\Big)}{\psi_{ik,n-1}^*\sum\limits_{l=1}^{K=2} \psi_{il,n-1}\Big(-\dfrac{\partial \mathcal{H}}{\partial \psi_{il,n-1}}+\lambda \psi_{il,n-1}^*\Big)}}, \quad i=1,\ldots,M; k=1,2.
\end{gather}
The final sonified output is then given by $\psi_{\text{sum},n}=\sum \limits_{i=1}^M [\psi_{i1,n}+\psi_{i2,n}]$. 
\par
\begin{figure}[!htbp]
	\begin{center}
		\includegraphics[scale=0.9,trim=2 2 2 2,clip]{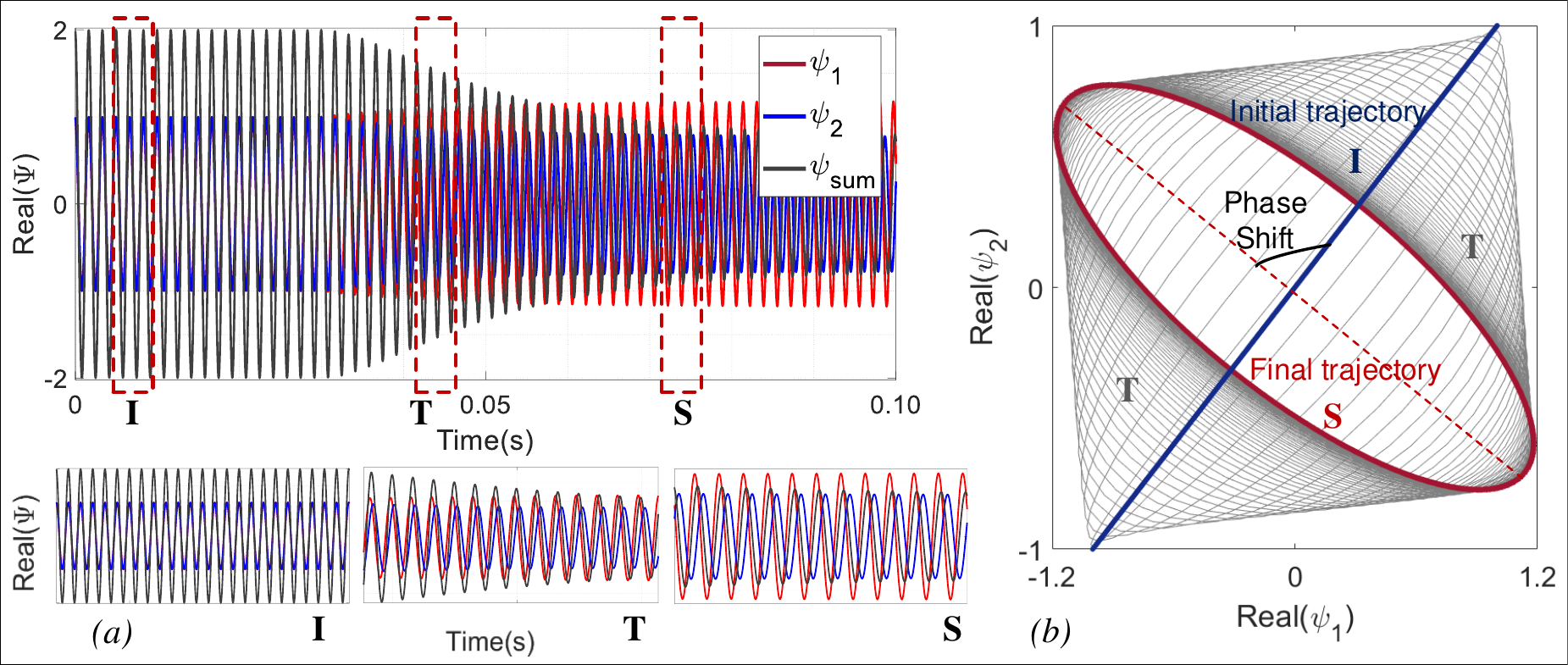}
		\caption[Time evolution and phase portraits]{\textbf{ Time evolution and phase portraits.}
			Time evolution and phase portrait for $\psi_1, \psi_2$ and $\Psi_{\text{sum}}$ corresponding to a 1D variant of the optimization problem $\mathcal{H}_2$ :(a) Time evolution of the waveforms, with zoomed-in versions of the initial(I), transient(T) and steady-state(S) regions respectively; (b) Phase portrait of $\psi_1$ vs. $\psi_2$. The oscillators go from an initial state(I) of the same amplitude and no phase difference, through a transient state(T) of varying amplitudes and varying phase difference, to a final steady-state(S) where they have constant amplitudes and a constant phase difference as well.}
		\label{fig5}
	\end{center}
\end{figure}

 \textbf{Time evolution and phase portrait of the sonified signal:} Fig \ref{fig5} shows the time evolutions and phase portrait for a one-dimensional problem ($M=1$), with $a=1, \mathbf{Q}=1, \mathbf{c}=0.8$ and $\gamma=2$. The sampling and natural/ baseline frequencies were considered to be $F_s=22 kHz$ and $\omega_n=600 Hz$ respectively, while the relative frequency was considered to be $\xi_{ik,n}=\xi_{k}=0$ for all the oscillators. In the sonification framework, $t = n \Delta t$, where $t$ seconds is the time duration corresponding to the $n-th$ time step, and each time step $\Delta t = (t+ \Delta t)- \Delta t=(n+1)\Delta t - n \Delta t = (1/Fs)$s. The simulation duration was 0.1s, with the optimization onset being at 0.03s for same initial amplitudes of $\psi_1$ and $\psi_2$, and zero initial phase difference between the waveforms. Fig \ref{fig5}(a) shows the time evolutions for the waveforms $\psi_{11}=\psi_1, \psi_{12}=\psi_2$ and $\psi_{\text{sum}}$, along with their zoomed-in views during the initial (I), transient (T) and steady-state (S) stages respectively. Fig \ref{fig5}(b) shows the phase portrait between the real parts of $\psi_1$ and $\psi_2$. It can be seen that the two waveforms evolve over time from an initial state (I) of same amplitude and zero phase difference, through a transient phase (T) of varying amplitudes and phase difference, to a final steady state where $\psi_1$ and $\psi_2$ have constant values of amplitudes, and a constant phase difference between them. This indicates in a non-zero phase shift of the final trajectory of $\psi_{\text{sum}}(t)$ from the initial trajectory.
\par
\begin{figure}[!htbp]
	\begin{center}
		\includegraphics[scale=0.6,trim=2 2 2 2,clip]{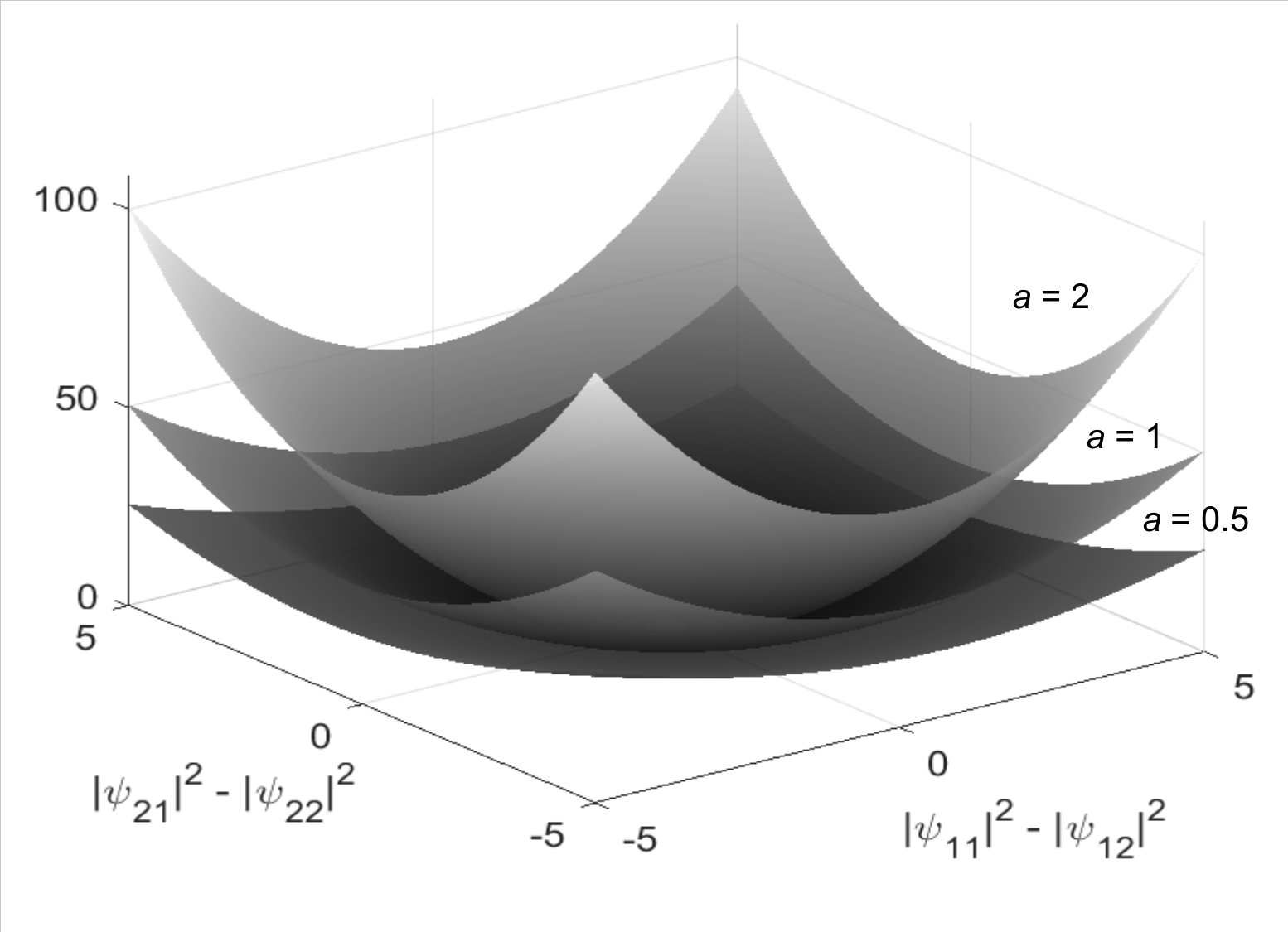}
		\caption[Quadratic optimization surface]{\textbf{ Quadratic optimization surface.}
			Optimization surface for the 2D variant of the problem $\mathcal{H}_2$ for different values of the parameter $a$.}
		\label{fig6}
	\end{center}
\end{figure}

\begin{figure}[!htbp]
	\begin{center}
		\includegraphics[scale=0.6,trim=2 2 2 2,clip]{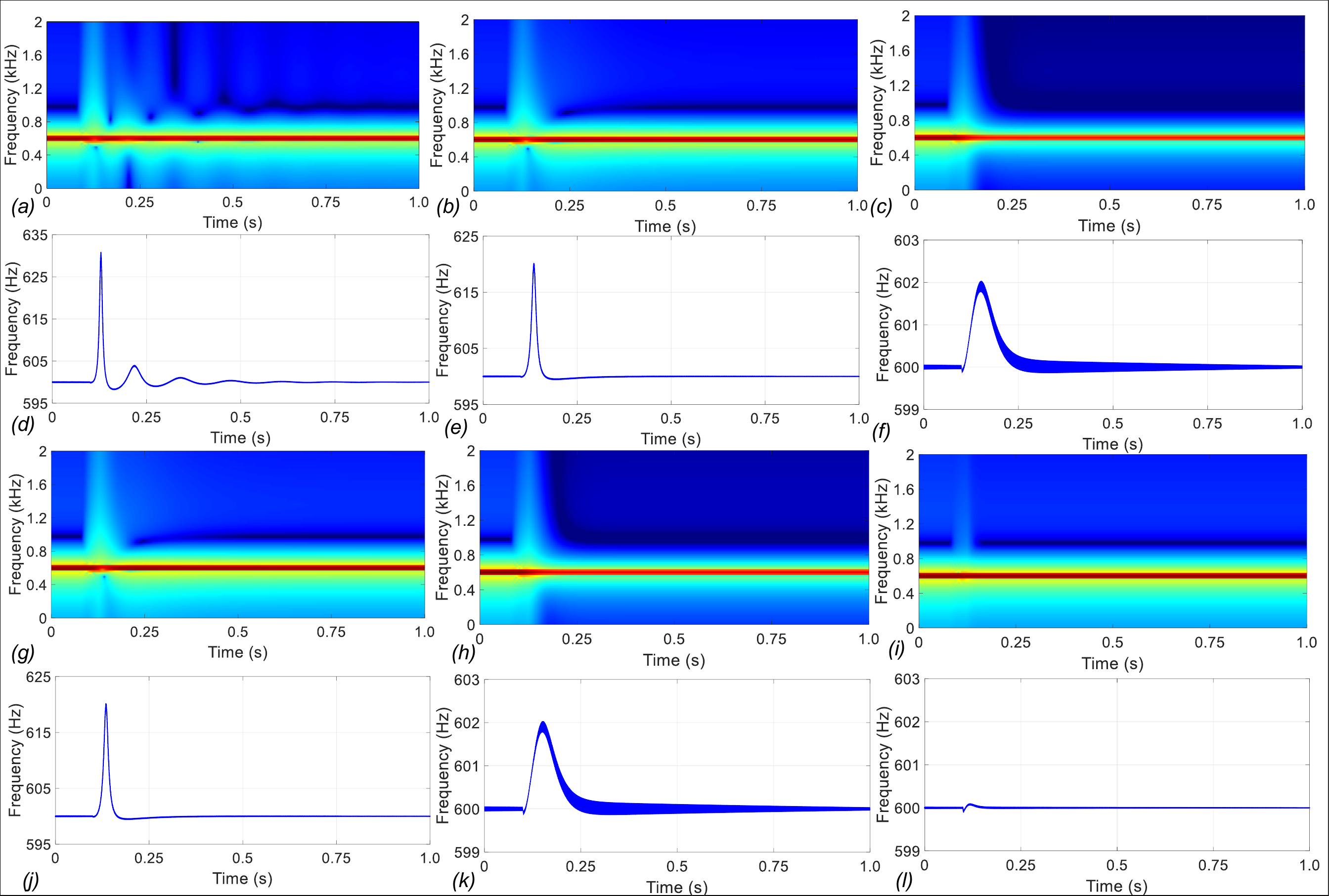}
		\caption[Effect of $\mathcal{H}$ and $\gamma$]{\textbf{ Effect of $\mathcal{H}$ and $\gamma$.}
			Variation of $\Psi_{\text{sum}}(t)$ with different values of curvature ($a$) and total energy available ($\gamma$) for a 2D variant of the optimization problem $\mathcal{H}_2$: (a)-(c) show the spectrograms;  (d)-(f) show the zoomed-in versions of the instantaneous frequency shift computed using Hilbert transform, for $a=0.5,1$ and $2$ respectively, and $\gamma=1$; (g)-(l) show similar plots for $\gamma=2.$}
		\label{fig7}
	\end{center}
\end{figure}

 \textbf{Effect of} $\mathcal{H}$ \textbf{and} $\gamma$ \textbf{on the sonified signal: }Next, we investigate the performance of the model for different values of curvature of the optimization surface as shown in Fig \ref{fig6}, as well as different levels of total energy available to the system at any instant of time. Fig \ref{fig7} demonstrates the model behavior for a two-dimensional variant of Example 2 (i.e., $M=2$). As before, we consider all the four oscillators in this case, namely, $\psi_{11},\psi_{12},\psi_{21},\psi_{22}$ to have the same value of natural frequency $\omega=600$ Hz and relative frequency $\xi_{ik}=\xi_{k}=0$. Simulations were carried out with $\mathbf{Q}=\mathbf{I}_2$ and $\mathbf{c}=0.8 \mathbf{1}_2$. Figs \ref{fig7}(a)-(c) show the spectrograms for the waveform $\Psi_{\text{sum}}(t)$, with $a=0.5,1$ and $2$ respectively, and $\gamma=1$. In all cases, the spectrogram was computed using a 1024-point Short Time Fourier Transform (STFT) with a sliding Kaiser window of size 1024 and steepness parameter 5, with an overlap size of 1023. Figs \ref{fig7}(d)-(f) show the zoomed-in versions of the instantaneous frequency shift computed using Hilbert transform. Figs \ref{fig7}(g)-(l) show similar results for   the same set of values for $a$ and a higher value of $\gamma=2$. The simulation duration is 1.0s, and the optimization process starts 0.1s after the start of the simulation. From the figures, we can conclude the following: (i) the duration of the transient phase as well as the magnitude of frequency shift during the same decreases with an increase in the value of $a$, (i.e., a steeper curve implies a faster optimization process and hence smaller values of frequency deviation); (ii) higher value of the total energy $\gamma$ leads to shorter transient phases with smaller frequency drifts. Thus, the complexity of the optimization problem is encoded in the final phase shift of the output signal from its initial phase.  
\par
\begin{figure}[!htbp]
	\begin{center}
		\includegraphics[scale=0.6,trim=2 2 2 2,clip]{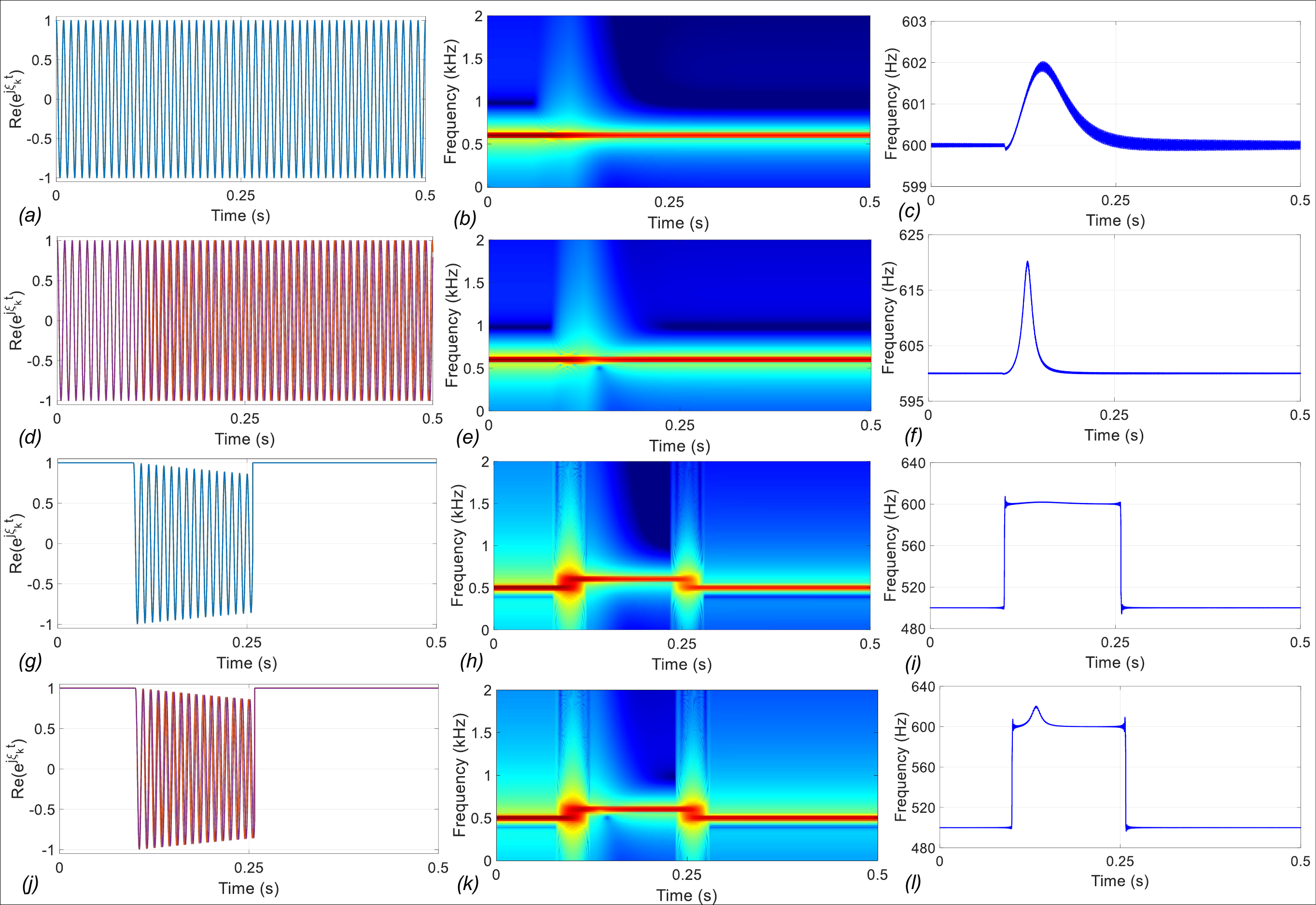}
		\caption[Effect of frequency perturbations induced by $\mathcal{H}$]{\textbf{ Effect of frequency perturbations induced by $\mathcal{H}$.}
			Variation of  $\Psi_{\text{total}}(t)$ for a 2D variant of the optimization problem $\mathcal{H}_2$, for different choices of the relative frequency $\xi_{k}(t)$: (i) \textbf{Case 1:}(a) $\xi_{k}$ is constant $\forall k,t$, (b) spectrogram of the output waveform, (c) instantaneous frequency shift using Hilbert transform ; (ii) \textbf{Case 2:} (d) value of $\xi_{k}(t)$ depends on the convergence rate of each variable $\psi_{k}(t)$, (e) spectrogram of the output waveform, (f)instantaneous frequency shift using Hilbert transform; (iii) \textbf{Case 3:} (g) $\xi_{k}(t)$ exists only during the optimization stage and is an exponentially decaying function which is the same for all $\psi_{k}$, (h) spectrogram of the output waveform, (i)instantaneous frequency shift ; (iv) \textbf{Case 4:} (g) $\xi_{k}(t)$ exists only during the optimization stage and is an exponentially decaying function which depends on the convergence rate of each variable $\psi_{k}$, (h) spectrogram of the output waveform, (i)instantaneous frequency shift. }
		\label{fig8}
	\end{center}
\end{figure}

 \textbf{Effect of cost function-induced frequency perturbations on the sonified signal: }Next, we examine the effect of variations in the relative frequencies on the system's performance. Fig \ref{fig8} shows the spectrogram and instantaneous frequency plots for different choices of time evolution of the relative frequency $\xi_{k}(t) \enskip \forall k$, for the optimization problem in Example 2 and same parameter settings as used in the simulations in Fig \ref{fig7}. The baseline frequency was chosen to be $\omega_n=100$ Hz for all the oscillators. Additionally, the total simulation duration and the onset of optimization were 1s and 0.2s, respectively, in all of the following experiments. 
\newline \textbf{\textit{Case 1 (Constant relative frequency)}}: For the first set of simulations, the relative frequencies are chosen to be $\xi_{k}=\xi=500$ Hz for all the oscillators, and no amplitude modulations or frequency modulations were added to the relative frequency trajectories. The plot of $\text{Re}(\exp(j\xi_{k}t))$, which is a unit amplitude sinusoid with a constant frequency for all the oscillators in this case, is shown in Fig \ref{fig8}(a). Figs \ref{fig8}(b) and (c) show the corresponding spectrogram and instantaneous frequency shift (computed using Hilbert transform) of the superimposed waveform $\psi_{\text{sum}}(t)$. 
\newline \textbf{\textit{Case 2 (Relative frequency modulation)}}: Figs \ref{fig8}(d)-(f) show similar results for the case when each relative frequency trajectory $\xi_{k}$ is modulated by the corresponding value of $\sigma_{k}$, where $\sigma_{k} \rightarrow 1$ in steady state, i.e., $\xi_{k,n}=\xi \sigma_{k,n}$. This type of frequency modulation thus incorporates information about the optimization process in the output signal, and leads to faster convergence and larger frequency deviation compared to the output in Case 1.
\newline \textbf{\textit{Case 3 (Amplitude decay and constant relative frequency)}}: In this case, a varying amplitude term constant for each oscillator was added to the relative frequency term, and no frequency modulation was added. The multiplicative term in the update equation is thus replaced by $\exp(j\xi_{k,n} \Delta t) \leftarrow \exp((-\rho+j\xi) \Delta t)$, where $\rho=1$ was chosen for the experiment.  Figs \ref{fig8}(g)-(i) show the relative frequency modulations, spectrogram, and frequency deviation respectively for this case.
\newline \textbf{\textit{Case 4 (Amplitude decay and frequency modulation)}}: In the last set of experiments, both the amplitude and frequency of the relative frequency trajectories were modulated for all the oscillators, i.e., $\exp(j\xi_{k,n}\Delta t) \leftarrow \exp((-\rho+j\xi \sigma_{k,n-1}) \Delta t)$. Figs \ref{fig8}(j)-(l) show the relative frequency modulations, spectrogram, and frequency deviation respectively for this case. In both Cases 3 and 4, the relative frequencies were applied only during the transient phase of the simulation, and set to zero both during the initial stage as well as after convergence to a steady-state. It can be seen that the frequency shift during the transient phase in Case 4 is significantly higher than all the other stages.

\par Based on the above experiments, we can infer the following:
\begin{itemize}
	\item The framework can encode information about the optimization problem's complexity and the total energy available to the network.
	\item Different encoding strategies can be employed by exploiting the relative frequencies of the oscillators.
	\item Frequency selectors/tuners can also be implemented by assigning sufficiently lower levels of energy to the oscillator network, such that only the oscillator with the maximum energy content is sustained.
\end{itemize}  

\subsection{Effect of Psychoacoustics on Sonified Output}
\label{sec_sonification_methods}

\par On the psychoacoustics side, we can incorporate both temporal and spatial attributes, as well as different choices of the sampling frequency, frequency mapping schemes, bandwidth, etc. Table \ref{tab_psychoacoustics} presents a summary of the various psychoacoustic parameters at our disposal. A temporal constraint is imposed on the sonified output by default because of the inherent power normalization provided by the growth transform framework. The temporal constraint thus imparts an automatic gain control feature \cite{lyon1990automatic} on the sonified signal. Since the output signal is inherently complex, spatial constraints can be imposed on the model using the left and right channels of the audio for the real and imaginary parts respectively, or vice versa. Finally, different sonification strategies can be adopted by considering different mapping schemes for both the baseline frequencies $\omega_i$'s, and the relative frequencies $\xi_{k}$'s. In a sense, the sonification process can be thought of as projecting high-dimensional data into a low-dimensional basis space created by predetermined frequency trajectories. We will present here three different sonification strategies: (a) using frequencies equally spaced on the Bark-scale, (b) creating chords based on a musical scale of choice, and (c) extracting dominant frequency trajectories from a chosen musical piece and mapping the basis set of frequencies to these trajectories. For generating human recognizable auditory signatures, all the baseline and relative frequencies should lie within the range of human perception, i.e., 20 Hz-20 kHz. Furthermore, the largest frequency assigned to a variable should be less than the Nyquist frequency to avoid aliasing.
\begin{table}[!htbp] 
	\renewcommand{\thetable}{\arabic{table}}
	\centering
	\caption{Psychoacoustic Parameters}
	\label{tab_psychoacoustics}
	\renewcommand{\arraystretch}{2}
	\begin{tabular}{|l|l|} \hline
		\centering
		\textbf{Constraint} & \textbf{Parameters}\\
		\hline
		Temporal & Power normalization \\ 
		\hline
		Spatial & Binaural hearing \\  
		\hline
		Frequency & Sampling frequency, frequency mapping, bandwidth \\ 
		\hline
		
	\end{tabular}
\end{table}

\par Some of the desirable characteristics of a candidate sonification strategy are as follows:
\begin{enumerate}
	\item Different data distributions can be encoded by different sound signatures, depending on the underlying optimization task.
	\item The complexity of the dataset or the underlying optimization problem affects the output sound signature. 
	\item For time-varying data, drift in the data distribution over time would lead to a drift in the sonified signal as well.
\end{enumerate}
\par For example, if we consider clustering as the underlying optimization problem, where the number of allowable clusters ($K$) is fixed apriori, then we would ideally want the sonified output to give an indication of (a) the instantaneous cluster densities, (b) the orientation of the clusters and (c) the time it takes for the optimization problem to converge to the optimal cluster assignments.
\subsubsection{Bark scale-based sonification}
\label{subsec_bark}

\par This method involves mapping the relative/baseline frequencies to equally spaced frequencies on the Bark scale. Any number of frequency trajectories can be selected if masking is acceptable in the end application. However, if we want to eliminate masking effects from the output sonified signal, the Bark scale frequencies should be chosen in a way such that the critical bands around these frequencies do not overlap with each other. Since the Bark scale has 24 critical bands, this implies that the number of frequencies in the basis set is limited to a maximum of 24 if we want to avoid masking. Additionally, the sonification module should be designed so that each frequency trajectory remains within its critical band on the Bark scale, even during the transient phase. This method's advantage is that changes in each frequency trajectory (i.e., each sonified variable) can be discerned unambiguously since there is no mixing of trajectories throughout the duration of sonification.

\subsubsection{Musical chord-based sonification}
\label{subsec_chord}

\par This approach is similar to the Bark scale-based approach, with the relative (or baseline) frequencies being mapped to a predetermined musical scale, e.g., the equally-tempered Western scale. Depending on the user requirements, we may create a chord by choosing notes in a single octave as the basis set. We can also select notes over multiple octaves, associating a different timbre to each, giving the impression of multiple different instruments being played. For example, we can map the frequencies to every other note in a diatonic scale such that they form a triad (e.g., the notes C, E, and G form the C major triad in the equally-tempered scale around $A_4 = 440$ Hz). Similarly, four or more random notes from the same octave create a generic chord. The method also allows for the creation of arpeggios (or a broken chord being played multiple times in succession) by suitably defining periodic repetitions of the set of chords. Overlapping of the frequency trajectories corresponding to different sonification variables may or may not occur, depending on (a) the pairwise distances between the frequency trajectories and (b) the maximum extent of frequency perturbation caused by the sonification module. Additionally,  since humans are more adept at recognizing time-varying audio signatures than static tones, a small, slowly varying sinusoidal variation may be added on top of the original frequency trajectories. The amplitudes and frequencies of these sinusoidal variations may be either (a) kept constant, or made to vary based on (b) the convergence properties of the optimization problem, or on (c) the instantaneous statistical properties of the sonification variables.
\subsubsection{Sonification using an existing musical piece}
\label{subsec_musical}

\par
\begin{figure}[!htbp]
	\begin{center}
		\includegraphics[scale=0.65,trim=2 2 2 2,clip]{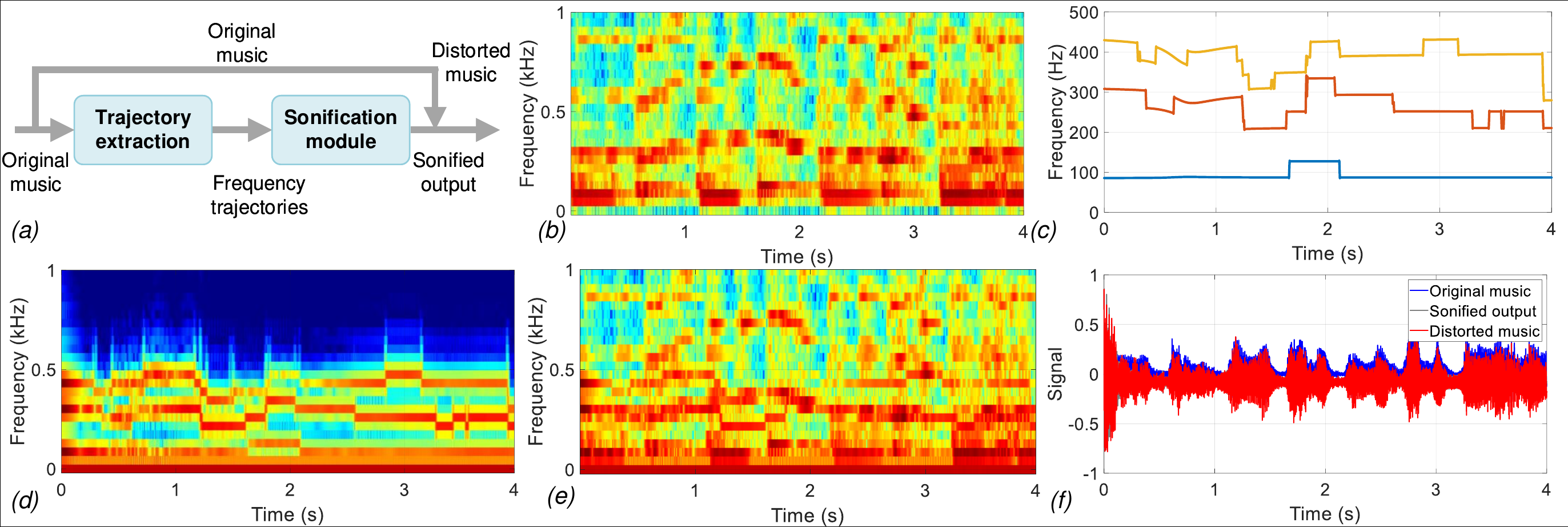}
		\caption[Musical composition based sonification technique]{\textbf{ Musical composition based sonification technique.}
			(a) Illustration of the sonification process in this technique; (b) Spectrogram of an original musical piece; (d) Raw frequency trajectories extracted from the music; (d) Spectrogram of the sonified output; (e) Spectrogram of the distorted music, obtained by superimposing the sonified output signal on the original music signal; and (f) Time evolution of the original musical composition, the sonified output, and the distorted music signal.}
		\label{fig9}
	\end{center}
\end{figure}

 This method of sonification involves extracting a certain number of dominant frequency trajectories (depending on the desired size $K$ of the basis set) from an existing musical piece. Additionally, the following steps need to be taken for extracting the dominant frequency trajectories from the chosen musical piece:
\begin{enumerate}
	\item Typically, a musical piece may be much longer in duration than the simulation duration and has a much higher sampling rate (44 kHz). Thus we would need to extract a segment of the entire composition and compress its time scale to ensure a proper mapping.
	\item Next, we analyze the spectrogram of the time-scaled musical sample and extract $K$ dominant frequency components in each time window of the spectrogram, based on the power content of each in that window.
	\item Finally, we carry out upsampling and interpolation to form continuous frequency trajectories that represent $K$ most dominant components of the musical composition.
\end{enumerate}
The sonification variables are then mapped to these frequency trajectories. The trajectories undergo perturbations in the transient phase during the sonification process, depending on the dataset complexity and the optimization problem being solved. The baseline frequency perturbations may remain unmodified or can be made a function of the convergence properties only. They can also be made to depend on certain statistical properties of the dataset or the optimization variables. Finally, for better interpretability, the sonification module's output can be treated as a ``noise" signal and superimposed on the original musical composition. The degree of deviation of this superimposed signal from the original composition thus encodes the dataset's complexity and the optimization process. Fig \ref{fig9} shows an overview of the musical composition-based sonification technique for a sample optimization problem, considering a basis set of size 3.

\section{Experiments on Synthetic Datasets}
\label{sec_syn_exp}
\par As a case study, we will consider a data clustering problem. The goal is to sonify the data by mapping each cluster to a particular tone or frequency trajectory, with the amplitude of each trajectory being proportional to the cluster density. Since the focus of the paper is on advocating a new sonification framework and not improving the efficiency of the particular clustering algorithm chosen, for the sake of simplicity, we will consider a similarity-based probabilistic clustering approach proposed in \cite{bulo2010probabilistic}. This involves solving a non-negative matrix factorization problem that minimizes the distance between a similarity matrix computed pairwise between the data points, and the actual likelihoods of the different data points to be clustered together in space. Consider a $D-$dimensional dataset $\mathbf{X} \in \mathbb{R}^{M \times D}$ and a similarity matrix $\mathbf{W} \in \mathbb{R}_+^{M \times M}$ computed using a pairwise distance metric between the data points. Then the following optimization problem assigns to the $i-$th data point a $K-$dimensional vector $\mathbf{p}_i$ consists of probability values of the data point of belonging to each of $K$ possible clusters:  

\begin{gather}
	\underset{\mathbf{P} \in \mathbb{R}_+^{M \times K},\beta \in \mathbb{R}_+} {\text{minimize}} \quad  \mathcal{H}(\mathbf{P}, \beta) = \lvert \lvert \mathbf{W}-\beta \mathbf{P}^T\mathbf{P} \rvert \rvert^2_2 \nonumber \label{eq_clustering1}\\
	\textit{s.t.}\quad \sum_{k=1}^K p_{ik}=1 \enskip \forall i=1, \ldots, M
\end{gather}

\begin{figure}[!htbp]
	\begin{center}
		\includegraphics[scale=0.7,trim=2 2 2 2 ,clip]{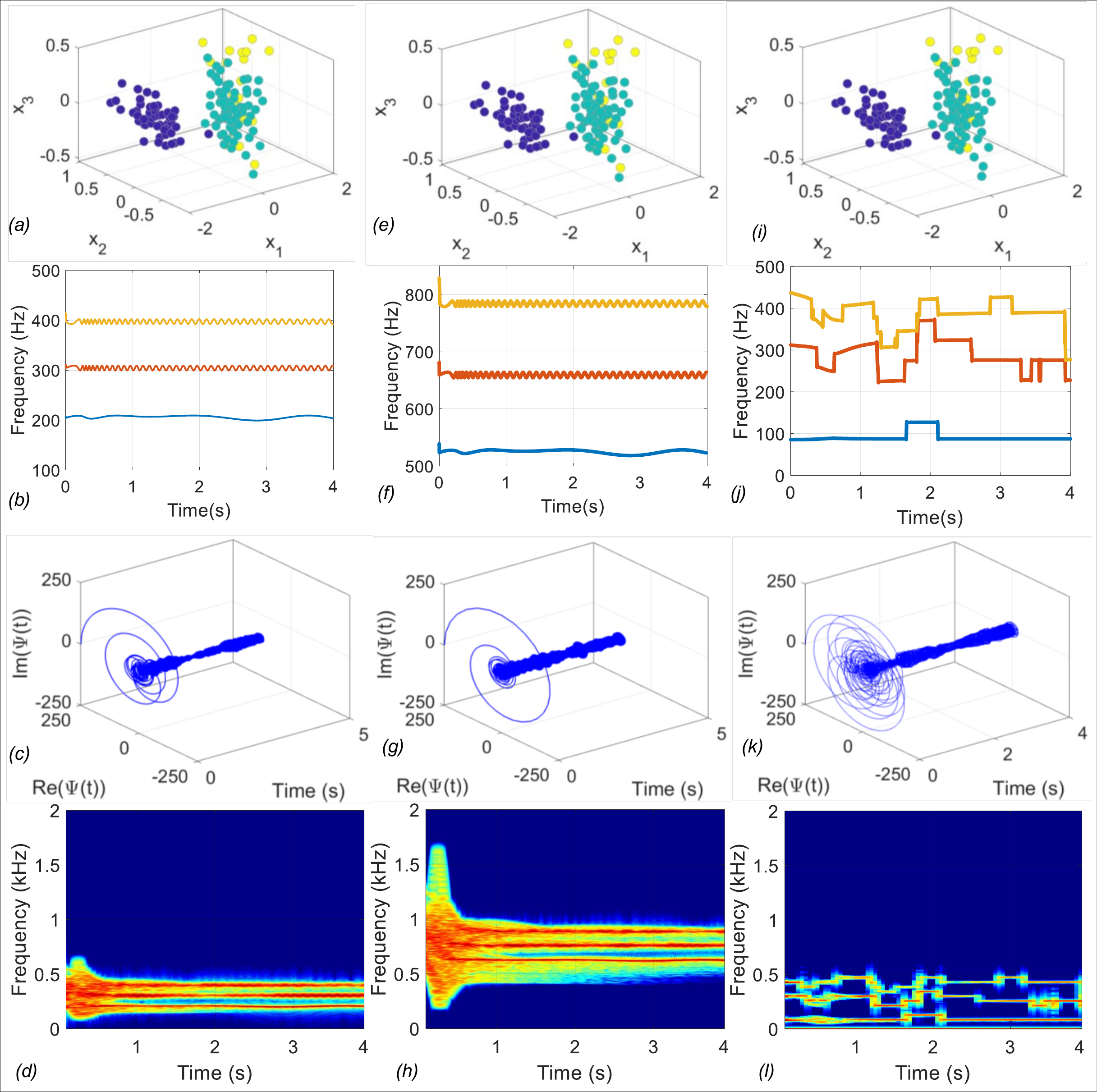}
		\caption[Different sonification strategies for clustering the `Iris' dataset]{\textbf{ Different sonification strategies for clustering the `Iris' dataset.}
			(a)-(d) show the PCA (labeled according to cluster assignments), frequency trajectories, time evolution, and spectrogram of the sonified output for the bark scale-based sonification; (e)-(h) show similar plots for the musical scale based sonification; and (i)-(l) show the results for the musical composition based sonification.}
		\label{fig10}
	\end{center}
\end{figure}

\begin{figure}[!htbp]
	\begin{center}
		\includegraphics[scale=0.7,trim=2 2 2 2,clip]{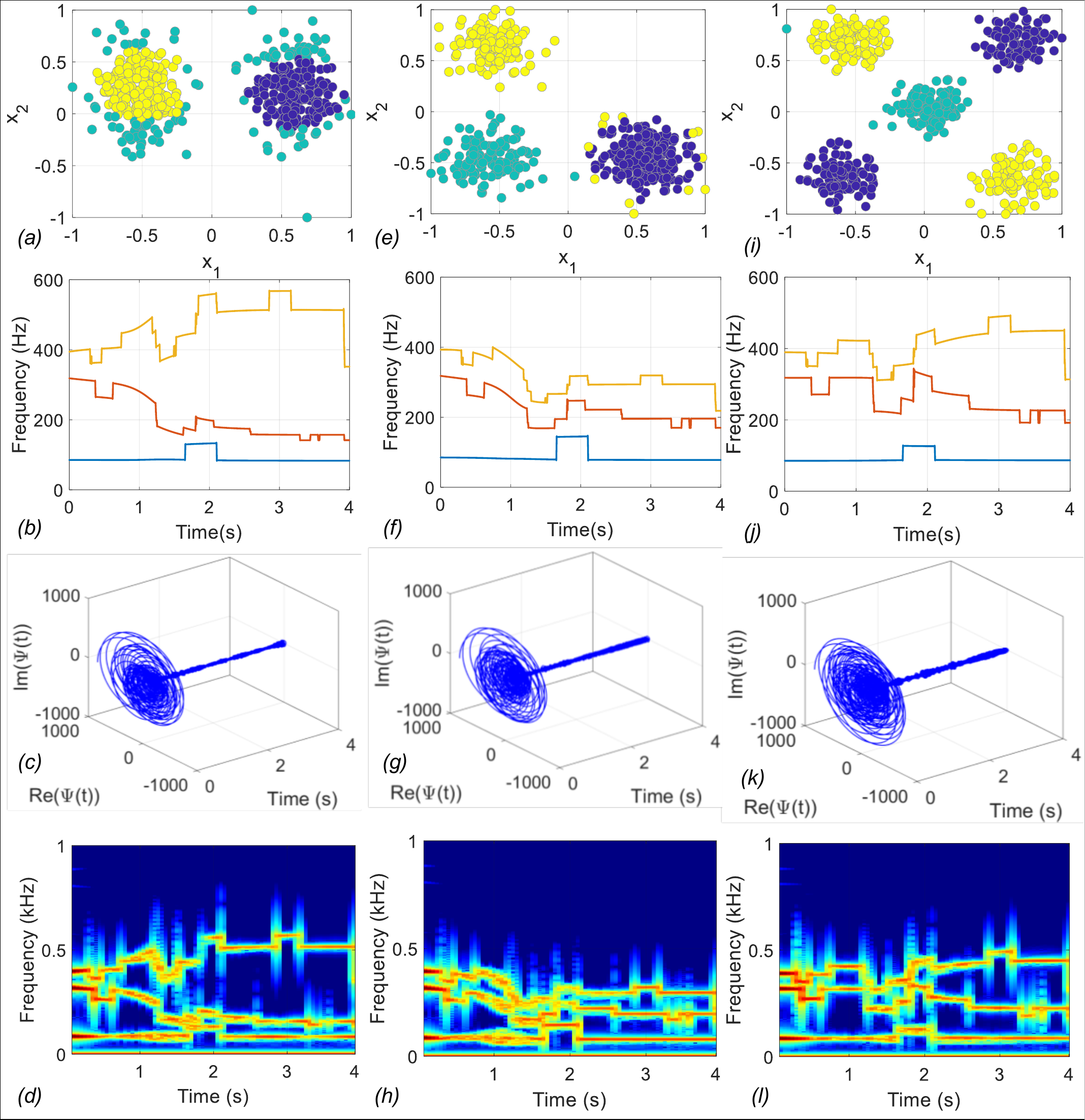}
		\caption[Musical composition based sonification on synthetic datasets for a basis set size of 3]{\textbf{ Musical composition based sonification on synthetic datasets with same number of points ($M=500$) but different number of clusters, with a basis set size of 3.}
			(a)-(d) show the scatter plot (colored according to the cluster assignments), frequency trajectories, time evolution, and spectrogram of the sonified output respectively for Dataset I; (e)-(h) show similar results for Dataset II; and (i)-(l) show the results for Dataset III.}
		\label{fig11}
	\end{center}
\end{figure}

\begin{figure}[!htbp]
	\begin{center}
		\includegraphics[scale=0.7,trim=2 2 2 2,clip]{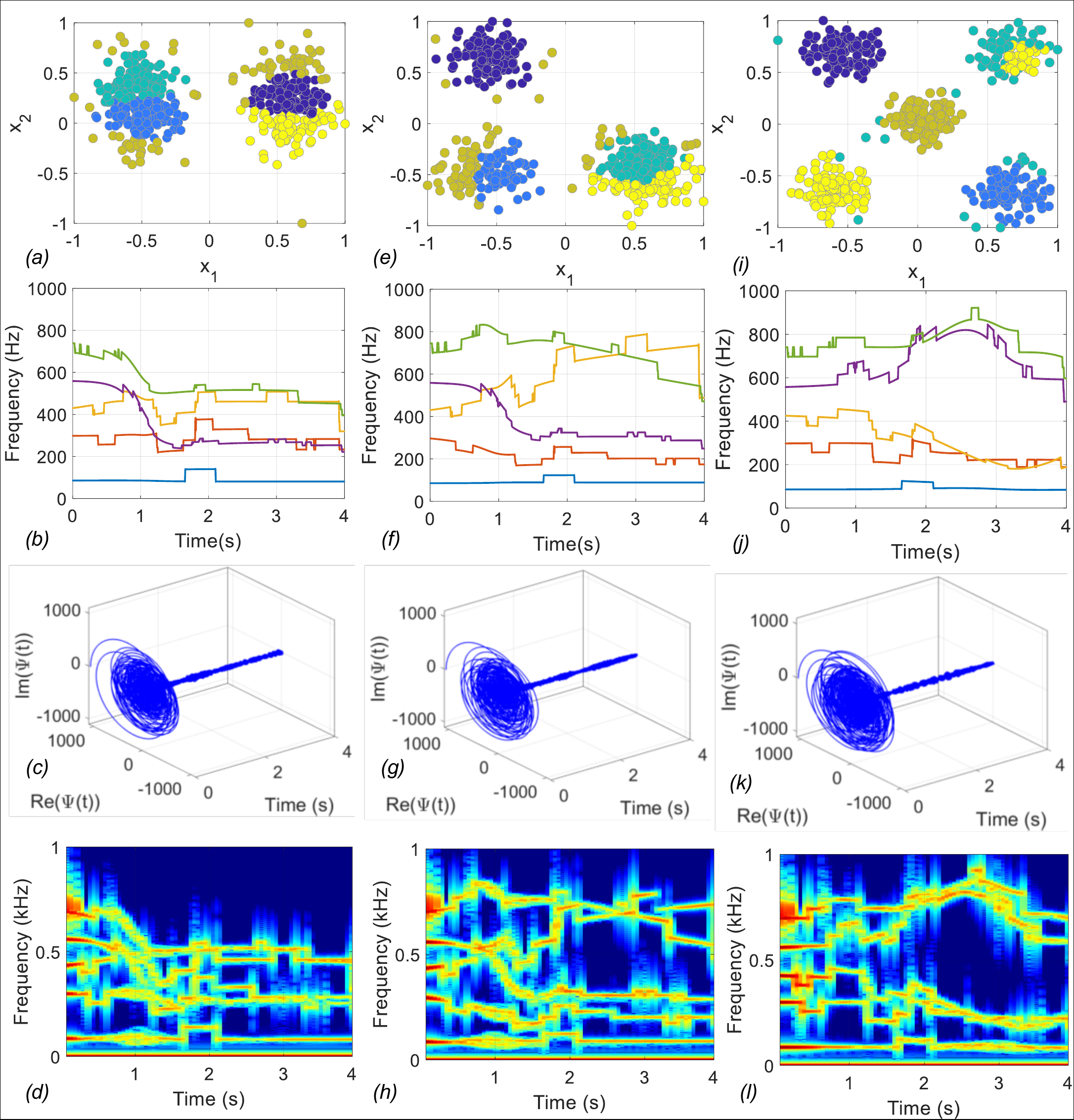}
		\caption[Musical composition based sonification on synthetic datasets for a basis set size of 5]{\textbf{Musical composition based sonification on synthetic datasets with same number of points ($M=500$) but different number of clusters, with a basis set size of 5.}
			(a)-(d) show the scatter plot (colored according to the cluster assignments), frequency trajectories, time evolution, and spectrogram of the sonified output respectively for Dataset I; (e)-(h) show similar results for Dataset II; and (i)-(l) show the results for Dataset III.}
		\label{fig12}
	\end{center}
\end{figure}

\begin{figure}[!htbp]
	\begin{center}
		\includegraphics[scale=0.6,trim=2 2 2 2,clip]{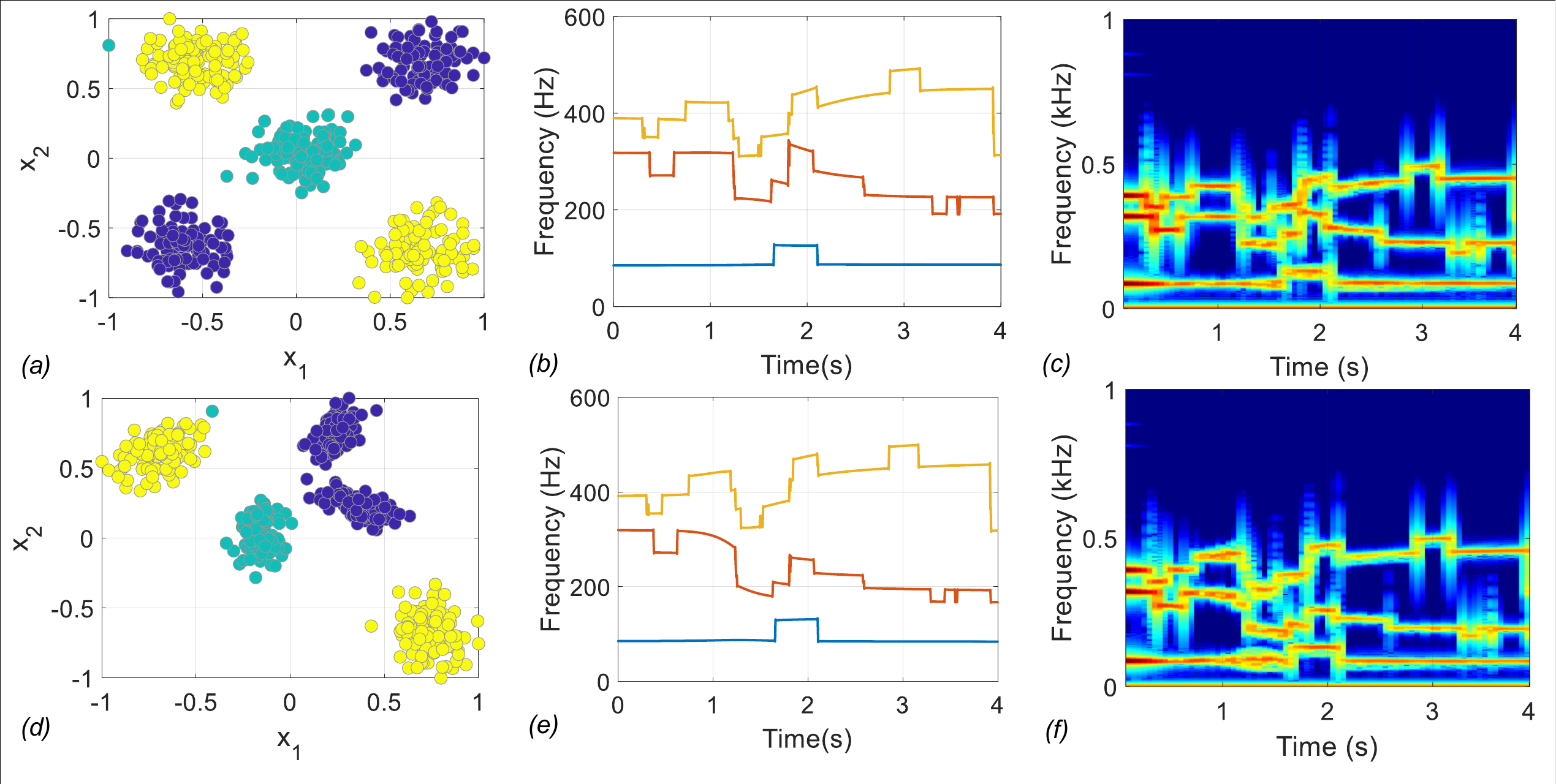}
		\caption[Musical composition based sonification on synthetic datasets for different cluster geometries]{\textbf{ Musical composition based sonification considering synthetic datasets with same number of points ($M=500$) and same number of clusters ($K=5$), but with different cluster geometries, considering a basis set of size 3.}
			(a)-(c) show the scatter plot (colored according to the cluster assignments), frequency trajectories, and spectrogram of the sonified signal respectively for Dataset III; and (d)-(f) show similar results for Dataset IV.}
		\label{fig13}
	\end{center}
\end{figure}

Thus, $p_{ik} \in \mathbb{R}_+$ denotes the probability of the $i$th data point of belonging to the $k$th cluster, and $\beta$ denotes a scaling factor such that $\beta \mathbf{p}_i^T \mathbf{p}_j$ represents the true likelihood of the $i$th and $j-$th data points to be clustered together. In this approach, each $w_{ij}$ is assumed to be normally distributed about its corresponding true likelihood $\beta \mathbf{p}_i^T \mathbf{p}_j$ with a constant variance. Here, the similarity matrix $\mathbf{W}$ is chosen to be the RBF kernel computed pairwise between the data points by mapping them to a high-dimensional space. 
We can use the procedure outlined in Table \ref{son_algo1} by applying the mapping $p_{ik}=\lvert \psi_{ik} \rvert ^2, \enskip \psi_{ik} \in \mathbb{C}$. Sonification of the clustering problem can then be achieved in the following manner:
\begin{itemize}
	\item We assign the same baseline frequency to all the subgroups (individual data points in this case), i.e., $\omega_i=\omega \enskip \forall i$.
	\item Each cluster is assigned to a particular relative frequency trajectory according to the sonification strategy chosen, i.e., $\xi_{ik}=\xi_k \enskip \forall i$. Depending on the sonification strategy, the unperturbed version of the relative frequencies $\xi_k(t)$'s may be chosen as follows:
	\newline \textit{(a) Bark scale based:} Each of the $K$ clusters is mapped to a distinct frequency on the Bark scale (with or without masking effects, depending on the frequency spacing). A slow sinusoidal variation may be added about each original frequency trajectory depending on the instantaneous cluster density, as well as the convergence characteristics of the optimization problem. The relative frequencies are thus modulated over time governed by the following equation:
	\begin{gather}
		\xi_k(t) \leftarrow \xi_k(t)[1+a_k \sin(2\pi b_k \Delta f_k(t) t)]s_k, \label{eq_sonifystrategy1} \\ 
		\text{where} \quad  \Delta f_k(t) = K \dfrac{c_{k}(t)-c_k(0)}{c_k(0)}, \enskip
		s_k = \dfrac{1}{N}\sum\limits_{i=1}^M \sigma_{ik}. \nonumber
	\end{gather}
	$a_k, b_k \in \mathbb{R}_+$ are constants, $c_k(0)=M/K$ is the initial cluster density (assuming the data points to be uniformly distributed among the clusters), and $c_k(t)=\sum\limits_{i=1}^M \lvert \psi_{ik} \rvert ^2$ is the instantaneous cluster density of the $k-$th cluster.
	\newline \textit{(b) Musical chord-based:} Each of the $K$ clusters is mapped to a distinct frequency on a chosen musical scale. In this case, too, a slow sinusoidal variation may be added about each original frequency trajectory depending on the instantaneous cluster density and the convergence characteristics of the optimization problem.  The evolution equations for the $\xi_k(t)$'s are similar to those used for the Bark scale-based method. 
	\newline \textit{(c) Using an existing musical composition:} Each of the $K$ clusters is mapped to a distinct frequency trajectory extracted from a musical composition. The instantaneous statistical properties of each cluster and the convergence properties of the optimization variables may be used as a scaling factor, to enhance the perturbations caused by the growth transform optimization framework. In this case, the evolution equations may have a sinusoidal variation as in the previous two approaches, or maybe of the following form:
	\begin{equation}
		\xi_k(t) \leftarrow \xi_k(t)[1+a_k \Delta f_k(t)]s_k
		\label{eq_sonifystrategy2}, 
	\end{equation}
	where $a_k, \Delta f_k(t)$ and $s_k$ have the same definitions as before. This is because the original frequency trajectories vary over time, and hence an additional sinusoidal perturbation for recognizing the audio signature in the steady state might not be necessary.
	\item The update equation for $\beta$ is given by:
	\begin{gather}
		\beta \leftarrow \dfrac{\Tr(\mathbf{WP}^T\mathbf{P})}{\lvert \lvert \mathbf{P}^T\mathbf{P} \rvert \rvert ^2},
	\end{gather}
	while those of the complex waveforms $\psi_{ik}$'s are obtained using the complex growth transform updates in Table \ref{son_algo1} (please see Appendix \ref{S1_appendix} for details).
	The time evolution of $\xi_k(t) \enskip \forall k$ occurs according to the update rules in (\ref{eq_sonifystrategy1}) or (\ref{eq_sonifystrategy2}), depending on the chosen sonification strategy.
	\item The final output of the sonification module is obtained by superimposing the waveforms of all the oscillators, i.e., $\psi_{\text{sum}}(t) = \sum \limits_{i=1}^M\sum \limits_{k=1}^K \psi_{ik}(t)$.
\end{itemize}
Next, we demonstrate the properties of the sonification technique when applied to the clustering problem for different sonification strategies, the actual number of clusters in the dataset, dataset complexity (i.e., cluster alignments), and the number of clusters assigned apriori (i.e., the size of the basis set of frequencies $K$).
\newline \textbf{Effect of different frequency mapping strategies:} Fig \ref{fig10} shows the results on the `Iris' dataset ($M=150$), considering a basis of size $K=3$, for all three types of sonification strategies discussed in Section \ref{sec_sonification_methods}. Fig \ref{fig10}(a) shows a PCA plot of the data for reference, where the data points have been colored according to their cluster assignments by the complex growth transform-based clustering algorithm. Fig \ref{fig10}(b) shows the frequency evolution plots of the relative frequency trajectories are chosen according to the Bark scale, with added sinusoidal variations, as discussed before. Fig \ref{fig10}(c) and (d) show the corresponding time evolution plot and the spectrogram respectively of the sonified output signal $\psi_{\text{sum}}(t)$. Figs \ref{fig10}(e)-(h) show similar results for the musical scale-based technique, while Figs \ref{fig10}(i)-(l) show those corresponding to the musical composition-based sonification.
\newline \textbf{Effect basis set size and number of clusters:} Fig \ref{fig11} shows the results of the sonification approach on synthetic Datasets I, II, and III, each containing $M=500$ data points, but the number of underlying clusters being 2,3 and 5 respectively. The synthetic datasets are generated using a Gaussian mixture model consisting of 2, 3 and 5 different clusters respectively, with a fixed cluster mean and variance associated with each cluster. We consider the basis set to consist of 3 frequency trajectories (i.e., $K=3$), and the musical composition-based sonification strategy for our experiments. Fig \ref{fig11}(a) shows a scatter plot of Dataset I, with the points color-coded according to their cluster assignment, Fig \ref{fig11}(b) shows the frequency evolutions, while Figs \ref{fig11}(c) and (d) show the time evolution and spectrogram of the sonified output, respectively. Fig \ref{fig11}(e)-(h) show the corresponding plots for Dataset II, while Fig \ref{fig11}(i)-(l) show those for Dataset III. Fig \ref{fig12} shows similar results on Datasets I, II, and III, for a basis set of frequencies of size $K=5$.
\newline \textbf{Effect of different cluster alignments:} Fig \ref{fig13} shows the results of the experiments on Datasets III and IV, where both have the same number of data points ($M=500$) and 5 underlying clusters, but differ in the cluster alignments and geometries (which represents different data complexities).

\section{Experiments on Real Datasets: Sonification of EEG recordings for Epileptic Seizure Detection}
\label{sec_real_exp}

\par In this section, we demonstrate how our framework can be applied for sonifying non-invasive electro-encephalogram (EEG) recordings for the detection of epileptic seizures. This is particularly useful in scenarios where trained physicians may not be readily available for analyzing the patterns in the EEG waveforms, as in the case of sub-clinical seizure onset. In instances of refractory epilepsy in particular, where patients are non-responsive to anti-epileptic drugs, neurostimulation or surgery is an option if the epileptogenic focus/foci can be identified \cite{parvizi2018detecting,scheid2021time}. Sonification can be useful in such scenarios since it usually implies faster feedback than visualization, leading to faster injection of the radiotracer for effective localization of the epileptogenic foci.

\par
\begin{figure}[!htbp]
	\begin{center}
		\includegraphics[scale=0.6]{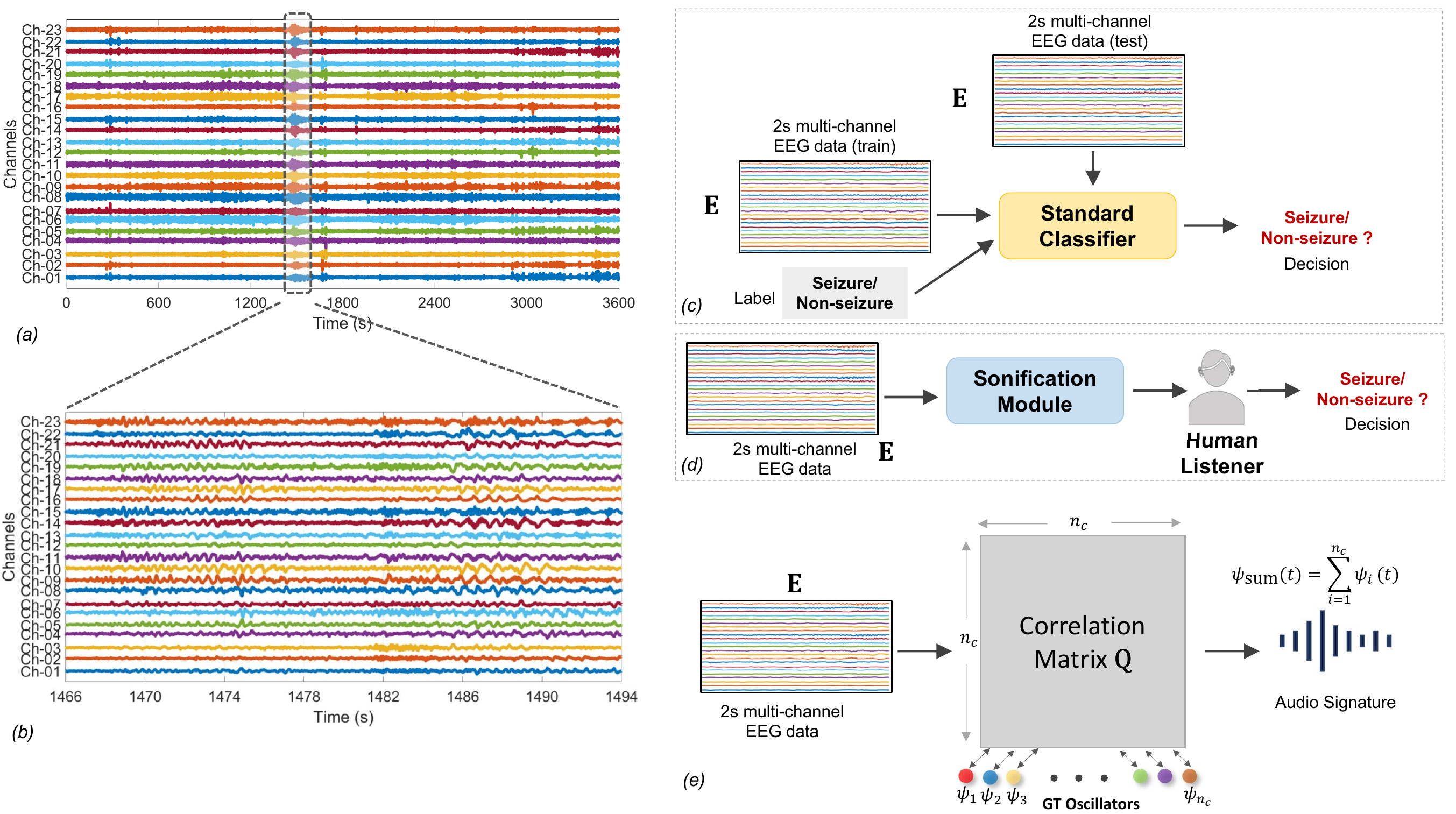}
		\caption[Sonification of EEG recordings]{\textbf{ Sonification of EEG recordings.}
			(a) 1 hour EEG recording with a single seizure event between 1467-1494s; (b) zoomed-in version of the seizure event; (c) An automatic seizure classification module which involves (i) a supervised learning phase using 2s windows of EEG recordings (represented by the matrix $\mathbf{E}$) along with the corresponding labels, and (ii) a testing phase where the presence of seizure is determined by using the trained model on a previously unseen 2s window; (d) Proposed approach, where features extracted from each 2s EEG window are passed directly through a sonification module and presented to a human listener, who takes the decision; and (e) Sonification module involves computing the pairwise connectivity matrix $\mathbf{Q}$ between all the $n_c$ EEG channels in the window. $\mathbf{Q}$ thus interconnects $n_c$ globally coupled growth transform oscillators, and the final sonified output is obtained by the superposition of all the oscillator outputs.}
		\label{fig14}
	\end{center}
\end{figure}

 Traditional approaches for seizure detection from EEG recordings include a sliding window-based feature extraction stage. This involves inferring network connectivities or frequency components of the EEG channels in each window \cite{bomela2020real}. The feature extraction stage is usually followed by a classification stage formed by support vector machine or neural network-based classifiers, and more recently, deeper architectures like CNNs \cite{zhou2018epileptic}. This is illustrated in Fig \ref{fig14}(c). On the other hand, sonification-based approaches for seizure detection usually adopt a parameter mapping approach where the voltage values of a single channel of the EEG signal are mapped to the auditory domain \cite{loui2014rapidly}. There exists a second approach for sonification of EEG signal. This involves mapping the output of the binary classifier into two distinct audio waveforms representing the presence or absence of seizure. While the first approach discards valuable information in terms of the network interactions during a seizure event, the second approach conveys the decision of the classifier using sound \cite{lin2018visualization}. It thus does not employ the end-user (clinician or caregiver) in the decision-making process.
\par In this section, we propose a novel sonification method that considers the interactions between all EEG channels to detect a seizure event, and creates a sonified output that enables the listener to make a decision. The proposed sonification module thus acts as a feature extractor in this case, and the task of detecting a seizure event still lies with the human-in-the-loop. We applied the sonification strategy to scalp EEG recordings of pediatric patients provided by the CHB-MIT database collected at the Children's Hospital Boston \cite{shoeb2010application,goldberger2000physiobank}. The recordings were collected from 23 channels placed according to the International 10-20 system of electrode positions, and all signals were sampled at 256 Hz. The original dataset contains EEG recordings from 23 patients (5 males aged 3-22 years, 17 females aged 1.5-19 years). Only the recordings that had one or more instances of the seizure (as marked by domain experts in the dataset) were used for our purpose. 

\par
\begin{figure}[!htbp]
	\begin{center}
		\includegraphics[scale=0.6]{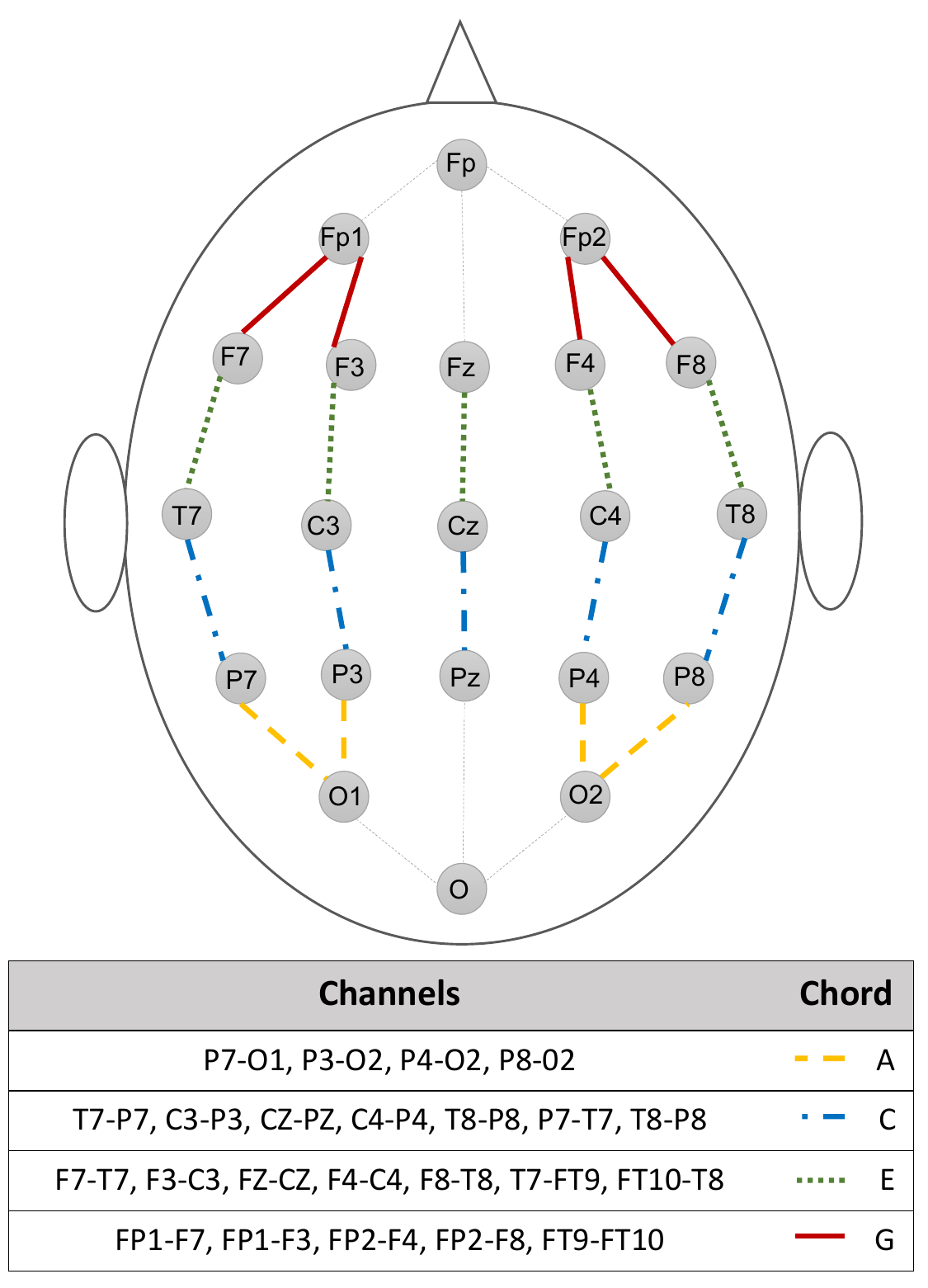}
		\caption[Relative frequency assignment to oscillators for EEG sonification]{ \textbf{Illustration of relative frequency assignment to oscillators.} Relative frequencies of oscillators were assigned to notes A, C, E, or G of the equally-tempered musical scale around A=440 Hz. Frequencies are assigned based on the electrode placement of the corresponding channel.}
		\label{fig15}
	\end{center}
\end{figure}

 The first step in the sonification process involves (i) removing the baseline, (ii) re-referencing the electrodes to the average potential, and finally (iii) applying a bandwidth filter with a passband of 0.5-50 Hz for noise reduction. All of these operations were performed using the publicly available EEGLAB toolbox \cite{delorme2004eeglab}. The EEG waveforms were then normalized channel-wise in the range $[-1,+1]$. Fig \ref{fig14}(a) shows a sample EEG recording with a seizure event lasting from 1467-1494s. Fig \ref{fig14}(b) shows a zoomed-in version of the seizure event. The EEG channel signals were then analyzed using a non-overlapping sliding window of 2s duration (denoted by the matrix $\mathbf{E}$). Note that 2s sliding windows have been used extensively in literature for epileptic seizure detection from EEG signals since they are short enough for retaining relevant information about the dynamics of the signal \cite{shoeb2010application}. Fig \ref{fig14}(c) shows the traditional supervised approach where the output of a classifier trained on features extracted from each of the 2s windows is converted to an audio waveform. In contrast, our proposed unsupervised approach (shown in Fig \ref{fig14}(d)) involves passing each such window through the sonification module to produce an audio signal. The output signal is finally used by the user to decide whether the window belongs to a seizure or a non-seizure event. In this framework, for each 2s epoch, we solve a quadratic optimization problem of the following form for inferring the relationship between different EEG channels:
\begin{gather}
	\underset{\mathbf{p} \in D \subset \mathbb{R}_+^{n_c} } {\text{minimize}} \quad  \mathcal{H}(\mathbf{p}) = \mathbf{p}^T \mathbf{Qp} \label{eq_eeg1}\\
	\textit{s.t.}\quad \sum_{k=1}^{n_c} p_{k} = 1,\enskip  
	p_{k}\ge 0 \enskip \forall k=1,\ldots,n_c, \label{eq_eegconstraint}
\end{gather}
where $n_c=23$ is the number of EEG channels under consideration. $\mathbf{Q} \in \mathbb{R}^{n_c \times n_c}$ represents a measure of the pairwise connectivity between the EEG channels, and is computed as $\mathbf{Q}=f(\mathbf{EE}')$. Here, $f(\cdot):\mathbb{R}^{n_c \times n_c} \mapsto \mathbb{R}^{n_c \times n_c}$ represents an element-wise square-root operator, and $\mathbf{E}$ represents the EEG data over the 2s window. Fig \ref{fig14}(e) illustrates the proposed approach. We can apply the complex growth transforms to arrive at the optimal solution of Eq (\ref{eq_eeg1})-(\ref{eq_eegconstraint}). For the simulations presented in this section, the baseline frequency of all the oscillators is chosen to be $\omega_n=300$ Hz. The relative frequencies $\xi_{k,n}$'s of the oscillator variables $\psi_k$'s (where $p_k=\psi_k \psi_k^*, \enskip \psi_k \in \mathbb{C}$) are chosen according to the relative placements of the electrodes from which the corresponding channels are recorded. For example, oscillators corresponding to channels recorded from the frontal, temporal, parietal, and occipital lobes of the brain are can be mapped to the notes G, E, C, and A based on the equally-tempered Western musical scale around A = 440 Hz. This mapping scheme based on the electrode locations is illustrated in Fig \ref{fig15}. Similar mappings can also be achieved using the Bark scale-based and musical composition-based approaches. Note that higher relative frequencies are assigned to the channels from the frontal and parietal lobes, as these brain regions are affected more by epileptic seizure events in pediatric patients than in the other two regions \cite{mass2021}. Thus allocating higher frequencies to the oscillators/channels mapping to these lobes would result in greater perturbation in the corresponding  frequency bands. This would potentially lead to better discrimination between seizure and non-seizure events. The framework thus allows for incorporating spatial information about the electrode locations because of the proposed mapping strategy, in addition to the temporal information captured by the sonification process. Spatial information can also be embedded in the sonification process by utilizing a dual-channel audio representation instead of the single-channel audio representation discussed until now in the paper. Since the output sonified signal is inherently complex, we can use the left and right channels for outputting the real and imaginary part respectively of the signal, or vice versa.

\par
\begin{figure}[!htbp]
	\begin{center}
		\includegraphics[scale=0.38,trim=2 2 2 2,clip]{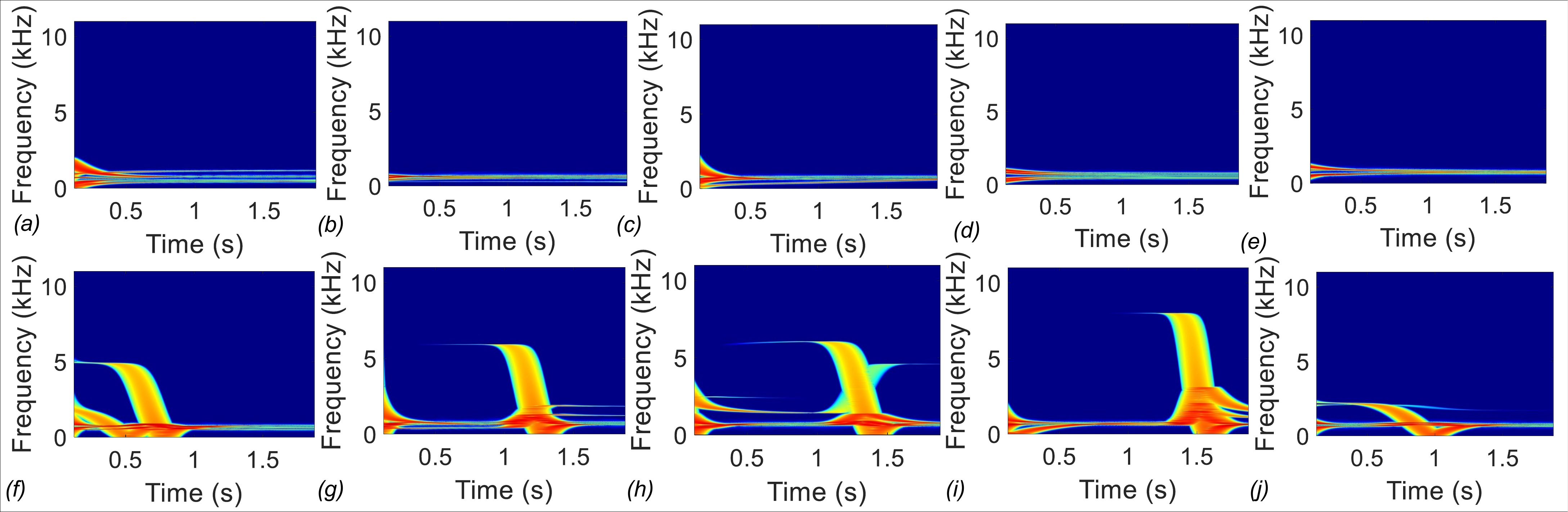}
		\caption[Sonification results on arbitrary 2s EEG windows for different recordings of the same patient]{{ \textbf{Sonification results on arbitrary 2s windows for 5 different recordings of Patient \#01} ( \#03, \#04, \#15, \#16 and \#18 respectively) .}
			(i) \textbf{Non-seizure events:} (a)-(e): Spectrograms for arbitrary non-seizure windows in each of the five recordings; (ii) \textbf{Seizure events:} (f)-(j): similar results for arbitrary seizure windows in each of the five recordings.}
		\label{fig16}
	\end{center}
\end{figure}

 Fig \ref{fig16} shows the results of applying our sonification technique to five different instances of seizure recordings (recording \#03, \#04, \#15, \#16, and \#18 respectively) for the same patient (Patient \#01). Figs \ref{fig16}(a)-(e) show spectrograms for 2s windows corresponding to non-seizure windows in each of the five recordings. 
Figs \ref{fig16}(f)-(j) show similar figures for 2s windows corresponding to seizure events in the same five recordings. Based on the above results, we can conclude that the spectrograms for the sonified windows corresponding to seizure events seem to have a higher level of entropy than those corresponding to non-seizure events.

\par
\begin{figure}[!htbp]
	\begin{center}
		\includegraphics[scale=0.38,trim=2 2 2 2,clip]{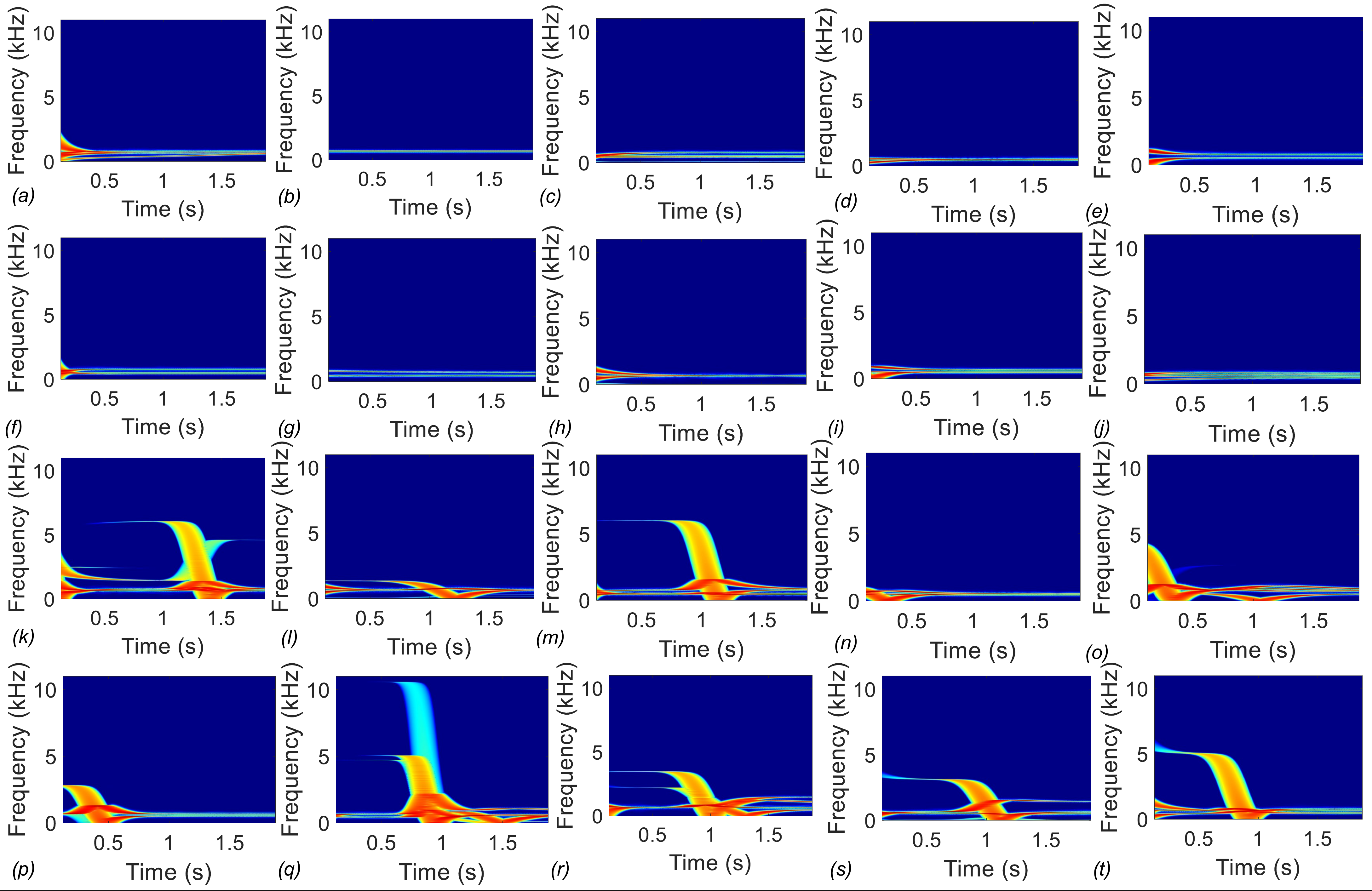}
		\caption[Sonification results on arbitrary 2s EEG windows for different patients]{{ \textbf{ Sonification results on arbitrary 2s windows for a single recording of each of 10 different patients} (Patient \#01, \#03 \#05, \#08, \#11, \#14, \#19, \#20, \#21 and \#22 respectively).}
			(i) \textbf{Non-seizure events}: (a)-(j): Spectrograms for arbitrary non-seizure windows in each of the ten patients;  (ii) \textbf{Seizure events} : (k)-(t): similar results for arbitrary seizure windows in each of the ten patients.}
		\label{fig17}
	\end{center}
\end{figure}

 Fig \ref{fig17}, on the other hand, presents a comparison of the results of the sonification process between seizure recordings across ten different patients (Patient \#01, \#03 \#05, \#08, \#11, \#14, \#19, \#20, \#21 and \#22 respectively). Figs \ref{fig17}(a)-(j) show the spectrograms for 2s windows corresponding to non-seizure events. Figs \ref{fig17}(k)-(t) show similar results for 2s windows corresponding to seizure events across the ten patients. The sonified signals corresonding to seizure vs non-seizure windows thus show consistent patterns not only across different recordings for the same patient, but also across patients. Our proposed sonification framework is thus capable of identifying both spatial and temporal patterns in the time-varying EEG signal, which can be potentially used for real-time detection of epileptic seizure events. 

\section{Discussion and Conclusions}

This paper presented a novel technique for the sonification of high-dimensional data that incorporates both learning and sonification stages into the same module. The end-user then uses the sonified audio output for making a decision. This is in contrast to existing sonification techniques that either involve (a) using a learning algorithm upstream that inherently controls the decision-making process and maps the outcome to an audio signature, or (b) directly mapping the underlying variables to different parameters of the sound wave, without accounting for the correlations and patterns in the data. At the core of the framework lies the complex growth transform dynamical system, which simultaneously utilizes the learning variables and psychoacoustic parameters defined by the user. Thus, our proposed sonification module outputs a binaural audio signature that can be used for human-in-the-loop decision-making. The output sonified signal encodes the high-dimensional space data, which might be particularly useful for low throughput systems. Additionally, the method can be utilized for solving a range of learning problems of varying dimensionality and provides several tunable parameters that can be customized to adapt different sonification strategies. Experiments on synthetic and real data show encouraging results, proving that the method can be used for real-time applications involving high-dimensional, temporally varying data. Future directions for this work involve explore using sonification in multimodal perceptualization tasks \cite{lemus2010sensory,recanzone2009interactions} with potential applications in visual rehabilitation \cite{jiang2015multisensory}, neuroprosthesis \cite{stiles2021multisensory}, and for cohesive perception in autonomous robots \cite{shiang2017multimodal}.

\section{Appendix}
\subsection{Nomenclature}
\label{S1_table}

Following are the notational conventions used in this paper:
\begin{table}[!htbp] 
	\renewcommand{\thetable}{\arabic{table}}
	\centering
	\caption{\bf Notations}
	\label{notations}
	\renewcommand{\arraystretch}{2}
	\begin{tabular}{|c|l|} \hline
		\centering
		\textbf{Variable} & \textbf{Definition} \\
		\hline
		$\mathbb{R}_+$ & One-dimensional positive real vector space \\ 
		\hline
		$\mathbb{R}^M$ & $M$-dimensional real vector space \\ 
		\hline
		$\mathbb{C}^M$ & $M$-dimensional complex vector space \\  
		\hline
		$\mathbf{P}$ & a matrix\\  
		\hline
		$\mathbf{p}$ & a vector\\  
		\hline		
		$\lvert z \rvert$ & magnitude of a complex variable $z$\\
		\hline
		$\text{Re}(z) $ & real part of a complex variable $z$\\
		\hline
		$\text{Im}(z) $ & imaginary part of a complex variable $z$\\
		\hline
		$z^*$ & complex conjugate of a complex variable $z$\\ 
		\hline 
		$z(t)$ & a continuous-time complex variable at time $t$ \\
		\hline
		$z_{n}$ & a discrete-time complex variable at the $n^{\text{th}}$ time step \\
		\hline
		$\odot$ & Hadamard product of two matrices\\
		\hline
	\end{tabular}
\end{table}

\subsection{Proof of Main Result}
\label{S1_appendix}

\textbf{Theorem 1:} \textit{Considering $\boldsymbol{\mathbf{\Psi}} \in D^C=\{\mathbf{\Psi}\in \mathbb{C}^{M\times K} :\sum\limits_{k=1}^{K} \lvert \psi_{ik}\rvert^2=1, \enskip \forall i=1,\ldots,M \}\subset \mathbb{C}^{M \times K}$, a time evolution of the form given below converges to limit cycles corresponding to the optimal point of a Lipschitz continuous objective function $\mathcal{H}(\boldsymbol{\Psi})$:}
\begin{align}
	\label{eq_thm1}
	{\psi_{ik,n}}\leftarrow \psi_{ik,n-1}[\cos(\theta_{i,n}) 
	+j\sin(\theta_{i,n})\sigma_{ik,n-1}(\boldsymbol{\Psi}_{n-1})] ,
\end{align}
\textit{where $\sigma_{ik} \rightarrow 1 \quad \forall i=1,\ldots,N, k=1,\ldots,M$, in steady state.}
\newline{Proof:}
Consider a constrained optimization problem of the following form : 
\begin{gather}
	\underset{\mathbf{P} \in D \subset \mathbb{R}_+^{M\times K} } {\text{minimize}} \quad  \mathcal{H}^{\prime}(\mathbf{P}) \label{eq_lemma1}\\
	\textit{s.t.}\quad \sum_{k=1}^Kp_{ik} = 1,\enskip \forall i=1,\ldots,M, \enskip p_{ik}\ge 0 \enskip \forall i,k  \label{eq_constraint}
\end{gather}
In \cite{chatterjee2018decentralized}, we used the Baum-Eagon inequality\cite{baum1968growth} to show that the optimal point of the optimization problem represented by Eqs~(\ref{eq_lemma1})-(\ref{eq_constraint}) for a generic Lipschitz continuous cost function $\mathcal{H}^{\prime}(\mathbf{P}) (\mathbf{P} \in D\subset \mathbb{R}^{M\times K}$)  corresponds to the steady-state solution of a multiplicative update-based discrete-time growth transform dynamical system model given by:
\begin{equation}
	p_{ik,n}\leftarrow (1-\alpha_{i,n})p_{ik,n-1}+\alpha_{i,n}p_{ik,n-1}g_{ik,n-1}(\mathbf{P}_{n-1}) ,
\end{equation}
where $g_{ik,n-1}(\mathbf{P}_{n-1})=\dfrac{(-\dfrac{\partial \mathcal{H}^{\prime}}{\partial p_{ik,n-1}}+\lambda)}{\sum\limits_{l=1}^K p_{il,n-1}(-\dfrac{\partial \mathcal{H}^{\prime}}{\partial p_{il,n-1}}+\lambda)}, \quad \forall i =1,\ldots,M$, and $0 \le \alpha_{i,n} \le 1 \quad \forall i$. The constant $\lambda \in \mathbb{R}_+$ is chosen to ensure that 
$\lvert -\dfrac{\partial \mathcal{H}^{\prime}(\mathbf{P})}{\partial p_{ik,n-1}}+\lambda \rvert > 0, \forall i,k$. Note that convergence to the optimal solution is guaranteed even though $\alpha_{i,n}$'s are time-varying, since this would still ensure the invariance of the manifold $D$. 
\par Taking $g_{ik,n-1}(\mathbf{P}_{n-1})= \sigma_{ik,n-1}^2(\mathbf{P}_{n-1}) \enskip \forall i,k$ and $\alpha_{i,n}=\sin^2(\theta_{i,n})$, we get:
\begin{align}
	p_{ik,n}\leftarrow   [\cos^2(\theta_{i,n})p_{ik,n-1}+\sin^2(\theta_{i,n})p_{ik,n-1}\sigma_{ik,n-1}^2(\mathbf{P}_{n-1})]
\end{align}
Representing $p_{ik,n}=\psi_{ik,n}\psi_{ik,n}^*, \enskip \psi_{ik,n} \in \mathbb{C}$, the update equations become:
\begin{align}
	\psi_{ik,n}\psi_{ik,n}^* 
	\leftarrow \psi_{ik,n-1}[\cos(\theta_{i,n})+j\sin(\theta_{i,n})\sigma_{ik,n-1}(\mathbf{P}_{n-1})]\nonumber \\
	\times \psi_{ik,n-1}^*[\cos(\theta_{i,n})+j\sin(\theta_{i,n})\sigma_{ik,n-1}(\mathbf{P}_{n-1})]^*
\end{align}
Considering $\mathcal{H}^{\prime}(\mathbf{P})=\mathcal{H}(\boldsymbol{\Psi})$ to be analytic in $D^C$, since by Wirtinger's calculus,
\begin{align}
	\dfrac{\partial \mathcal{H}}{\partial \psi_{ik,n-1}} = &\dfrac{\partial \mathcal{H}}{\partial \psi_{ik,n-1}\psi_{ik,n-1}^*}.\Bigg(\dfrac{\partial \psi_{ik,n-1}\psi_{ik,n-1}^*}{\partial \psi_{ik,n-1}}\Bigg)
	= & \dfrac{\partial \mathcal{H}}{\partial \psi_{ik,n-1}\psi_{ik,n-1}^*}.\psi_{ik,n-1}^*,
\end{align} 
we have 
\begin{gather}
	\sigma_{ik,n-1}(\boldsymbol{\Psi}_{n-1})=\sqrt{\dfrac{\Big(-\dfrac{\partial H}{\partial  \psi_{ik,n-1}}+\lambda \psi_{ik,n-1}^*\Big)}{\psi_{ik,n-1}^*\sum\limits_{l=1}^K \psi_{il,n-1}\Big(-\dfrac{\partial \mathcal{H}}{\partial \psi_{il,n-1}}+\lambda \psi_{il,n-1}^*\Big)}}
\end{gather}
The discrete time update equations for $\psi_{ik,n}$ is thus given by:
\begin{align}
	{\psi_{ik,n}} \leftarrow  \psi_{ik,n-1}[\cos(\theta_{i,n}) 
	+j\sin(\theta_{i,n})\sigma_{ik,n-1}(\boldsymbol{\Psi}_{n-1})],
	\label{eq_fin_discrete}
\end{align}
Now, let us define $\theta_{i,n}=\omega_{i,n}\Delta t$, where $\omega_{i,n}$ is the baseline frequency for the $i-$th group of oscillator variables at the $n-$th time step, and $\Delta t$ is a unit time step. Eq (\ref{eq_fin_discrete}) can then be rewritten as:
\begin{align}
	{\psi_{ik,n}} \leftarrow  \psi_{ik,n-1}[\cos(\omega_{i,n}\Delta t) 
	+j\sin(\omega_{i,n}\Delta t)\sigma_{ik,n-1}(\boldsymbol{\Psi}_{n-1})],
	\label{eq_fin_discrete2}
\end{align}
\newline \hspace*{\fill} \qeda
\newline \textbf{Theorem 2:} \textit{Different oscillation frequencies can be assigned to each element in the dynamical system represented by Eq (\ref{eq_thm1}) by adding an instantaneous relative phase term, i.e., the following dynamical system}
\begin{align}
	{\psi_{ik,n}} \leftarrow  \psi_{ik,n-1}[\cos(\omega_{i,n}\Delta t) 
	+j\sin(\omega_{i,n}\Delta t)\sigma_{ik,n-1}] 
	\exp(j\xi_{ik,n}\Delta t),
	\label{eq_thm2}
\end{align}
\textit{converges to the same solution as that attained by the one described by Eq (\ref{eq_thm1}).}
\newline Proof: Considering $\varphi_{ik,n}$ to be the instantaneous relative phase of the $ik-$th element with respect to an absolute reference at the $n-$th time instance, we have:
\begin{align}
	\psi_{ik,n} \leftarrow \psi_{ik,n-1}[\cos(\omega_{i,n}\Delta t) 
	+j\sin(\omega_{i,n}\Delta t)\sigma_{ik,n-1}] 
	\exp(j\varphi_{ik,n}) \label{eq_preunitary},
\end{align}
Let $\varphi_{ik,n}$ be the instantaneous relative phase of the $ik$-th oscillator variable. Then we can define $\varphi_{ik,n}=\xi_{ik,n} \Delta t$, such that $\xi_{ik,n}$ is the corresponding relative frequency. The addition of the relative frequency term only alters the oscillator dynamics without altering the instantaneous solution of the optimization problem. Additionally, we can incorporate information about the instantaneous oscillator dynamics in the relative phase by using $\varphi_{ik,n}=c_{ik}\sigma_{ik,n}\xi_{ik,n} \Delta t$, where $c_{ik}$'s are arbitrary scaling coefficients. Note that as before, this mapping only perturbs the dynamics, without impacting the true solution. Eq (\ref{eq_preunitary}) can be rewritten as
\begin{align}
	\psi_{ik,n} \leftarrow  \psi_{ik,n-1} \sqrt{\cos^2(\omega_{i,n}\Delta t)+\sin^2(\omega_{i,n}\Delta t)\sigma_{ik,n-1}^2} \nonumber \\
	\times \exp \Bigg(j \Big(\arctan(\sigma_{ik,n-1}\tan(\omega_{i,n}\Delta t)) + \xi_{ik,n} \Delta t \Big) \Bigg) \label{eq_preunitary2},
\end{align}
Finally, in the steady state, since $\sigma_{ik,n-1} \xrightarrow{n \rightarrow \infty} 1 \enskip \forall i,k $, and taking $\Delta t = \dfrac{1}{Fs}$ ($Fs$ is the sampling frequency), we have:
\begin{align}
	\psi_{ik,n} \leftarrow  \psi_{ik,n-1} \exp \Bigg(j(\omega_{i,n}+ \xi_{ik,n})\dfrac{1}{Fs}\Bigg) \label{eq_preunitary_steady}.
\end{align}
The output sonified signal in the steady-state is then obtained by the superposition of all the oscillator waveforms, i.e.,
\begin{equation}
	\psi_{\text{sum},n} = \sum \limits_{i=1}^M \sum \limits_{k=1}^K \psi_{ik,n} = \sum \limits_{i=1}^M \sum \limits_{k=1}^K \psi_{ik,n-1} \exp \Bigg(j(\omega_{i,n}+ \xi_{ik,n})\dfrac{1}{Fs}\Bigg)
\end{equation}
In particular, when there is a single global constraint involving all the variables, i.e., when $M=1$, then $\psi_{ik,n}=\psi_{k,n}, \omega_{i,n}=\omega_n$ and $\xi_{ik,n}=\xi_{k,n}$. The complex growth transform dynamical system for this special case can be written as:
\begin{align}
	\psi_{k,n}  & \leftarrow  \psi_{k,n-1} \exp \Bigg(j(\omega_{i,n}+ \xi_{ik,n})\dfrac{1}{Fs}\Bigg) \\
	\psi_{\text{sum},n} & = \sum \limits_{k=1}^K \psi_{k,n-1} \exp \Bigg(j(\omega_{n}+ \xi_{k,n})\dfrac{1}{Fs}\Bigg).
\end{align}
\newline \hspace*{\fill} \qeda
\newline \textbf{Theorem 3:} \textit{The complex growth transform dynamical system follows a nonlinear unitary transformation.}
\newline Defining

\begin{gather}
	\label{eq_unitary}
	\mathcal{U}_{n} = 
	\left(
	\begin{array}{ccccc}
		d_{11,n-1}                                    \\
		& \ddots             &   &  \\
		&               & d_{ik,n-1}                \\
		&  &   & \ddots            \\
		&               &   &   & d_{MK,n-1}
	\end{array}
	\right), \nonumber  \\
	\text{where} \enskip d_{ik,n}=[\cos(\omega_{i,n}\Delta t) 
	+j\sin(\omega_{i,n}\Delta t)\sigma_{ik,n-1}]\exp(j\xi_{ik,n}\Delta t),
\end{gather}

and using Eq (\ref{eq_preunitary}), we have: 		
\begin{equation}
	\boldsymbol{\Psi}_n \leftarrow \mathcal{U}_{n}(\boldsymbol{\Psi}_{n-1}) \odot \boldsymbol{\Psi}_{n-1}.
\end{equation}
Additionally, since $\sum \limits_{i=1}^M \sum \limits_{k=1}^K \lvert \psi_{ik,n} \rvert ^2 = \sum \limits_{i=1}^M \sum \limits_{k=1}^K \lvert \psi_{ik,n-1} \rvert ^2$, $\mathcal{U}_{n}: \mathbb{C}^{M \times K} \mapsto \mathbb{C}^{M \times K}$ can be thought of as an instantaneous nonlinear unitary operator \cite{schwartz1997nonlinear} that ensures that the signal energy is conserved over time.
\par For the special case of a single global constraint ($M=1$), we have
\begin{gather}
	\label{eq_unitary2}
	\mathcal{U}_{n} = 
	\left(
	\begin{array}{ccccc}
		d_{1,n}                                    \\
		\vdots            \\
		d_{k,n}                \\
		\vdots            \\
		d_{K,n}
	\end{array}
	\right), \quad \mathcal{U}_n :\mathbb{C}^K \mapsto \mathbb{C}^K, \nonumber  \\
	d_{k,n}=[\cos(\omega_{n}\Delta t) 
	+j\sin(\omega_{n}\Delta t)\sigma_{k,n-1}]\exp(j\xi_{k,n}\Delta t), \quad \forall k=1,\ldots,K \\
	\text{and} \enskip  \sum \limits_{k=1}^K \lvert \psi_{k,n} \rvert ^2 =  \sum \limits_{k=1}^K \lvert \psi_{k,n-1} \rvert ^2.
\end{gather}
\newline \hspace*{\fill} \qeda

\subsection{Mapping procedure}
\label{S2_appendix}

\par  Let us consider an optimization problem of the following generic form:
\begin{align}
	\underset{\{p_{i} \in \mathbb{R}_+\}} {\text{min}} \quad  \mathcal{H}^{\prime}(\{p_{i}\})  \nonumber  \\
	\textit{s.t.} \enskip \lvert p_{i} \rvert \le \gamma, \enskip \gamma \in \mathbb{R}_+ \enskip \forall i. 
\end{align}
Considering $p_{i}=p_{i}^+-p_{i}^- \enskip \forall i=1,\ldots,N$, where both $p_{i}^+,p_{i}^- \ge 0$. Since by triangle inequality, $\lvert p_{i} \rvert \le \lvert p_{i}^+ \rvert + \lvert p_{i}^- \rvert$, enforcing $p_{i}^+ + p_{i}^-=\gamma \enskip \forall i$ would automatically ensure $\lvert p_{i} \rvert \le 1 \enskip \forall i$. Thus we have,
\begin{align}
	\underset{\{p_{i}\}} {\text{argmin}} \quad \mathcal{H}^{\prime}(\{p_{i}\})&  \equiv \underset{\{p_{i}^+,p_{i}^-\}} {\text{argmin}} \quad \mathcal{H}^{\prime}(\{p_{i}^+,p_{i}^-\}) \label{eq_twovar}\\
	\textit{s.t.} \quad \lvert p_{i} \rvert \le \gamma, \enskip p_{i} \in \mathbb{R} & \quad \quad  \textit{s.t.} \quad p_{i}^+ + p_{i}^-=\gamma, \enskip p_{i}^+,p_{i}^- \ge 0  \nonumber 
\end{align}
\par Finally, we replace $p_{i}^+ \leftarrow \lvert \dfrac{p_{i}^+}{\gamma} \rvert, \enskip p_{i}^- \leftarrow \lvert \dfrac{p_{i}^-}{\gamma} \rvert$ to arrive at the following equivalent optimization problem over a probabilistic domain:
\begin{align}
	\underset{\{p_{i}^+,p_{i}^-\}} {\text{min}} \quad  \mathcal{H}^{\prime}(\{p_{i}^+,p_{i}^-\})  \nonumber  \\
	\textit{s.t.} \enskip p_{i}^+ + p_{i}^-=1,  \enskip \forall i. \label{eq_optform2}
\end{align}

The above problem, in turn, can be mapped to the complex growth transform-based sonification framework by considering $p_i^+ = \lvert \psi_{i1} \rvert ^ 2$, $p_i^- = \lvert \psi_{i2} \rvert ^ 2$, and following the procedure in \ref{S1_appendix}.

\bibliographystyle{IEEETran}
\bibliography{ref_Sonification}

\end{document}